\documentclass[12pt]{article}
\usepackage{fullpage, amsmath, amsthm, amsfonts, rotating}
\usepackage{hyperref}
\usepackage{nameref}
\usepackage{multirow}
\usepackage{lmodern,xcolor}

\usepackage{graphicx}
\usepackage{amssymb}
\usepackage{epstopdf}
\usepackage{natbib}   
\newcommand{\bmx}{\mbox{\boldmath $x$}}

\def \N{\mbox{N}}
\def \v{\mbox{var}}

\newcommand{\bmB}{\mbox{\boldmath $B$}}

\newcommand{\bms}{\mbox{\boldmath $s$}}
\newcommand{\bmS}{\mbox{\boldmath $S$}}

\newcommand{\bx}{\mbox{\boldmath $x$}}

\newcommand{\bbeta}{\mbox{\boldmath $\beta$}}
\newcommand{\bmbeta}{\mbox{\boldmath $\beta$}}

\newcommand{\bomega}{\mbox{\boldmath $\bomega$}}

\newcommand{\bc}{\begin{center}}
\newcommand{\ec}{\end{center}}

\newcommand{\bi}{\begin{itemize}}
\newcommand{\ei}{\end{itemize}}

\newcommand{\be}{\begin{enumerate}}
\newcommand{\ee}{\end{enumerate}}

\newcommand{\bs}{\begin{slide*}}
\newcommand{\es}{\end{slide*}}

\newcommand{\rT}{\mbox{\tiny{T}}}
\title{Small Area Estimation of Health Outcomes}
\author{Jon Wakefield$^{1,2}$,  Taylor Okonek$^1$ and Jon Pedersen$^3$\\ \\
$^1$Department of Biostatistics, University of Washington, Seattle, USA\\ $^2$Department of Statistics, University of Washington, Seattle, USA\\$^3$Fafo AVF, Oslo, Norway}

\begin{document}
\maketitle

\noindent
{\bf Abstract: }  Small area estimation (SAE) entails estimating characteristics of interest for domains, often geographical areas, in which there may be few or no samples available. SAE has a long history and a wide variety of methods have been suggested, from a bewildering range of philosophical standpoints. We describe design-based and model-based approaches and models that are specified at the area-level and at the unit-level, focusing on health applications and fully Bayesian spatial models.  The use of auxiliary information is a key ingredient for successful inference when response data are sparse and we discuss a number of approaches that allow the inclusion of covariate data. SAE for HIV prevalence, using data collected from a Demographic Health Survey in Malawi in 2015--2016, is used to illustrate a number of techniques. The potential use of SAE techniques for outcomes related to COVID-19 is discussed.

\vspace{.2in}

\section{Introduction}

Small area estimation (SAE) describes the endeavor of producing estimates of quantities of interest, such as means and totals, for domains (usually areas) which have sparse or non-existent response data. SAE is carried out in many fields including health, demography, agriculture, business, education, and environmental planning. In this article we focus on health outcomes, for which SAE aids in highlighting geographical disparities, and is broadly useful for health planning, resource allocation and budgeting. SAE allows fundamental questions such as: ``how many people in my area have condition X or need treatment Y".
Disease mapping which is traditionally based on a complete enumeration of disease cases, and differs from SAE which is typically based on a subset of individuals, selected via a survey, which may have a complex design. The standard reference on SAE is \cite{rao:molina:15}, while an excellent review is provided by \cite{pfefferman:13}. 

In survey sampling, inference has often focused on the {\it design-based} (or {\it randomization})
approach. This focus has carried over into SAE. This approach is quite distinct from the {\it model-based} approaches
that are the bread and butter of mainstream spatial 
statistics. Both inferential approaches are discussed in \cite{skinner:wakefield:17}. Design-based methods assess the frequentist properties of estimators, averaging over all possible samples that could have been drawn, under the specified sampling design. Under this paradigm, the values of the responses in the population are viewed as fixed rather than random. Model-based approaches can be either frequentist or Bayesian. If a model-based approach is taken, a hypothetical infinite population model is specified for the responses, which are now viewed as random variables. Modeling may be carried out within the design-based paradigm via model-assisted approaches \citep{sarndal:etal:92} in which a model is specified but desirable design-based properties are retained, even under model misspecification.  A cautious view \citep{lehtonen:veijanen:09} is that design-based (including model-assisted) inference may be reliable in situations when there are large or medium samples in areas, while if data are sparse, a model-based approach may be a necessity. In a companion article, \cite{datta:09} reviews, and is more enthusiastic toward, model-based approaches. Random effects modeling is popular under the model-based approach to SAE, and inference for these models is often frequentist through empirical Bayes estimation -- this is what we refer to as frequentist model-based inference, rather than the predictive approach described in \cite{valliant:etal:00}.

If a model-based approach is taken, a key element of model specification (and a source of contention) is determining how to account for the sampling design. Models may be specified at the {\it area-level} or at the {\it unit-level}. For the former, an important reference is \cite{fay:herriot:79}, which introduced the idea of modeling a weighted estimate of an area-level characteristic. For the latter, \cite{battese:etal:88} describe a nested error regression model at the level of the sampling unit.

Design-based inference from sample surveys aims to estimate finite population quantities using information about the sampling process. In a model-based framework, the finite population is itself a sample from a superpopulation a random process that can be described by some model. If, in an actual survey, the realized sample and the finite population can be described by the same model, then the sample design is ignorable. A simple random sample (SRS) is ignorable in this sense.  However, most real-life household surveys have non-ignorable designs, and most SAE models depend on assumptions. When the design is not ignorable, one needs to incorporate the design into the model.  Ideally, such incorporation  would include the relevant aspects of the design including design weighting, non-response corrections, and weight adjustments (see Section \ref{sec:notation}). While this information may be available, many aspects of the sampling frame (such as the locations of all clusters in a design with cluster sampling) are typically not available, at least not in sufficient detail to be useful.  For surveys such as the Demographic and Health Surveys (DHS), which are extensively carried out in low- and middle-income countries (LMIC), stratification, clustering and estimation weights will typically be available. For surveys in developed countries, little information may be available.

Prevalence mapping \citep{wakefield:20:prevalence}, is a name that has been given to the production of maps displaying the prevalence of health and demographic outcomes, and this endeavor clearly has large overlaps with SAE.  While SAE smoothing methods often use area-level models, prevalence mapping often uses model-based geostatistics (MBG) methods in which a continuous spatial model is specified. Examples of prevalence mapping using area-level SAE techniques include HIV prevalence \citep{gutreuter:etal:19} and  the under-5 mortality rate (U5MR) \citep{dwyer:etal:2014,mercer:etal:15,li:etal:19}.  Examples of prevalence  mapping with MBG include HIV prevalence  \citep{dwyer2019mapping}, malaria \citep{gething:etal:16}, U5MR \citep{golding:etal:17} and vaccination coverage \citep{utazi:etal:18}.

\begin{figure}[htbp]
    \centering
    \includegraphics[width=0.8\linewidth]{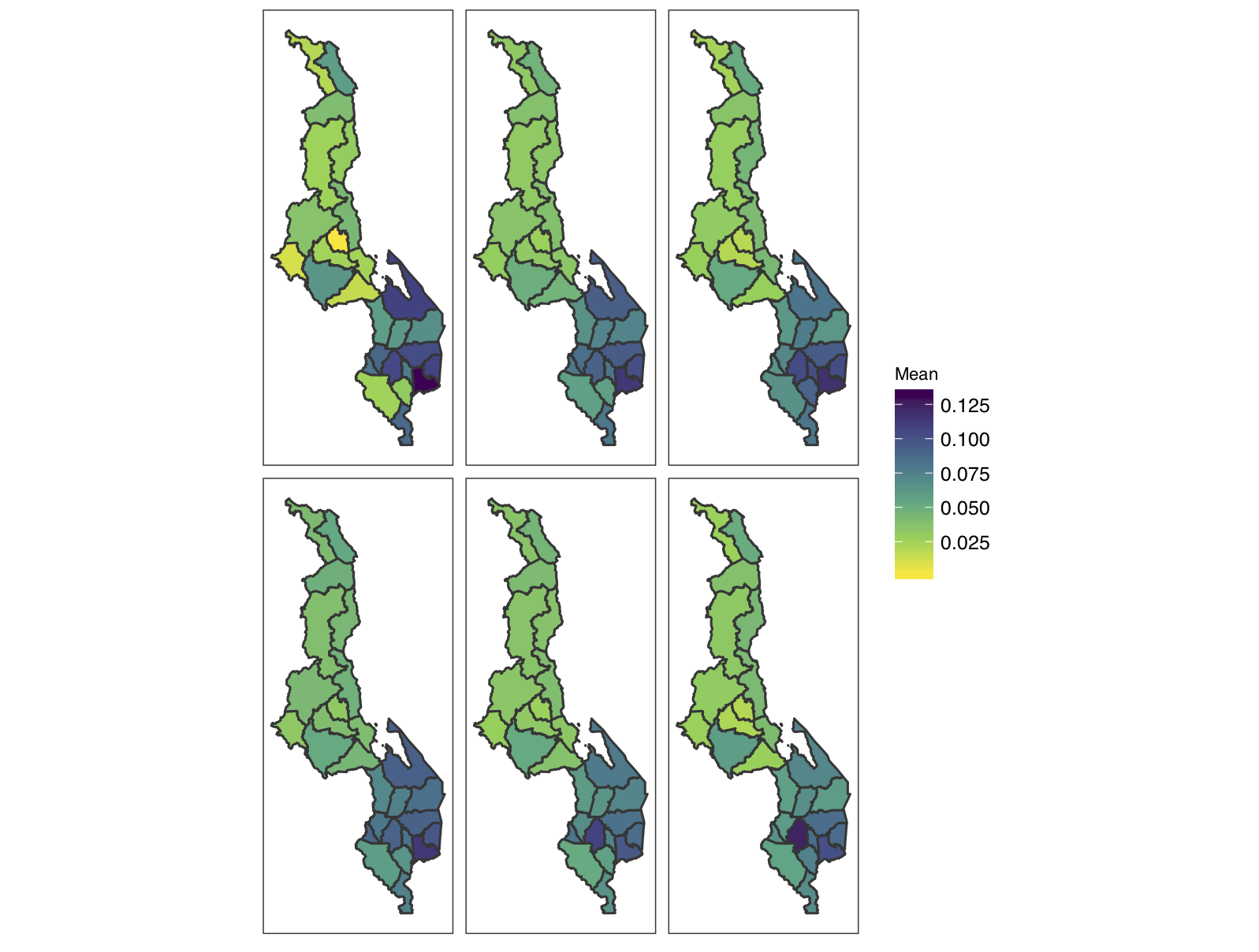}\\
    \caption{Estimates of HIV prevalence among females aged 15--29 in districts of Malawi in 2015--16. Top row estimates are from area-level models: direct estimates; smoothed direct estimates; smooth direct estimates with ANC HIV prevalence covariate. Bottom row estimates are from unit-level models: no urban/rural adjustment and no covariate; urban/rural adjusted only; urban/rural strata and ANC HIV prevalence covariate.}
    \label{fig:Malawi1}
\end{figure}

As a motivating example we consider SAE of HIV prevalence among females aged 15--29, in districts of Malawi, using data from the 2015--16 Malawi DHS. We will refer to the districts as admin-2 areas; in Malawi there are 3 admin-1 areas, 28 admin-2 areas and 243 admin-3 areas -- here we are using the GADM (database of global administrative areas) classification (\url{https://gadm.org/download_country_v3.html}). A two-stage stratified cluster sample was implemented, with the sampling clusters (enumeration areas) being stratified by district and urban/rural. The Malawi Population and Housing
Census (MPHC), conducted in Malawi in 2008 provided the sampling frame for the survey. The sample for the 2015--16 Malawi DHS was designed to
provide estimates of key indicators for the country as a whole, for urban and rural areas separately, and for
each of the 28 districts. The sampling frame contained 12,558 clusters and our analyses use data from 827 sampled clusters (the supplementary materials give more details). In the 2015--16 DHS survey for Malawi, 8,497 women in the age range 15--49 were eligible for testing, and 93\% of them were tested. HIV prevalence data was obtained from voluntarily taken blood samples from DHS survey respondents. Testing is anonymous, and as such, respondents cannot receive their test results directly from DHS. Instead, they are given educational materials and referrals to voluntary, free counseling and testing. Blood samples are collected on filter paper via finger pricks before being sent to a laboratory for testing \citep{malawi2016}.

Urban clusters were oversampled, relative to rural clusters (details are in the supplementary materials). We dropped the island district of Likoma, because it is spatially distinct (and quite culturally different from the nearest points on the mainland) and has a very small population.  In the top left panel of Figure \ref{fig:Malawi1}, we plot the weighted prevalence estimates at the district level and see great variation and what appears to be an increasing gradient in HIV prevalence from north to south. However, the uncertainty (shown in the supplementary materials) is relatively large, particularly in the south. Ignoring the survey design gives a national prevalence estimate (standard error) of 6.28\% (0.37\%) for females aged 15--29, while the weighted estimate is 6.18\% (0.48\%). The weighted estimate is smaller because the weights account for the urban oversampling (HIV is more prevalent in urban areas)  and the increased standard error is due to the clustering.


The structure of this article is as follows. In Section \ref{sec:notation}, notation and basic ideas are presented before traditional methods of direct and indirect estimation are described in Section \ref{sec:traditional}. We focus on spatial methods that have not been covered in recent SAE reviews in Section \ref{sec:spatial}, with both area-level and unit-level models being described. Models of each kind are applied to the HIV prevalence data in Section \ref{sec:HIV}.  The potential application of SAE methods to COVID-19 modeling is discussed in Section \ref{sec:COVID} and the paper concludes with a discussion in Section \ref{sec:discussion}.

\section{Notation and Overview}\label{sec:notation}

We will let $i$ index the areal units for which estimation is required, with $m$ areas in total. Assume there are $N_i$ individuals in the population with responses $y_{ik}$, $k=1,\dots,N_i$, in area $i$. 
We let $S_i$ represent the set of indices of the selected individuals in area $i$, with $n_i = |S_i|$ the number sampled in area $i$. From a design-based perspective $n_i$ may or may not be random, depending on the survey plan -- for example, if the small areas (domains) correspond to the strata in a stratified design (sometimes called {\it primary domains}) then the $n_i$ will be non-random. It is more typical for this not to be the case, and in this situation we have {\it secondary domains}. Under a model-based view, $n_i$ is conditioned upon and therefore fixed. We take as target of inference $m_i=\frac{1}{N_i} \sum_{k=1}^{N_i} y_{ik}$, the empirical mean of the response in the finite population. As examples of binary outcomes, interest may focus on the proportion of a population in an area that have a disease, have been vaccinated, or who are practicing social distancing (a much softer endpoint). This is a finite population characteristic, and such targets are common in survey sampling. Another common target is the total $ \sum_{k=1}^{N_i} y_{ik}$. In the model-based approach, the hypothetical  infinite population mean in area $i$ is denoted $\mu_i$. This can be interpreted as the mean of the distribution from which the finite population was (hypothetically) drawn.

We will focus on household surveys which are extremely popular, both in high-income countries (for example, in the United States, the National Crime Victimization Survey and the American Community Survey) and in LMIC (for example, the DHS, and Multiple Indicator Cluster Surveys).
Sampling of respondents for household surveys based on face-to-face interviewing is usually conducted via stratified multi-stage cluster sampling. A sample of primary sampling units (PSUs) are drawn  from what is typically an area-based sampling frame, such as enumeration areas in a population census. Most surveys employ stratification, usually by residence (urban and rural) and administrative units. Since stratification usually has very low cost, and somewhat reduces the variance of estimates, there is usually no reason to avoid stratification.  From the point of view of estimating totals across the whole sample, an allocation of sampling units to strata that is proportional to the population size is preferable. Optimal allocation \citep[Section 3.4.2]{lohr:19}, where allocation of sampling units is proportional to the standard deviation of the estimator and size of the stratum, is preferable if the variances of the target estimator in each stratum are known and there is a single target estimator. In the case of most household surveys, the aim is to provide estimates of many different population quantities that all have different variances. Optimal allocation is, therefore, not possible. Proportional allocation ensures that the variances will not be worse than if the sample was by SRS. However, many surveys are disproportionally stratified because of the need to report on particular domains with similar precision, such as reporting by urban and rural or individual provinces.

Nearly all surveys select PSUs by probability proportional to size (PPS), where the size measure is the number of households in each PSU as recorded in the sampling frame. A linear systematic PPS sample  \citep[Section 8]{murthy:rao:88} can be obtained if the sampling units are arranged contiguously along the real number line, each unit occupying a space equal to its size $ N_i$. Let $T_i = T_{i-1}+N_i$, $i=1,\dots,m$, with $T_0=0$ and the sampling interval be $t=T_m/n$. Then, select a random point $r$ from 1 to $t$ and
select unit $i$ if,
$$ T_{i-1} < r+ j t \leq T_i,\qquad j=0,1,\dots,n-1.
$$
If the size measure is constant for all sampling units, the sample becomes a linear systematic SRS. 

In a two-stage sample, which is the most common design, a list of households in each selected PSU is constructed, usually by mapping the cluster and listing every household by visiting every dwelling. A fixed number of households are then drawn using SRS or linear systematic SRS. The method has the benefit that the total sample size, both in total and in each cluster,  is fixed by design. The fixed sample sizes thus reduce variance and ease the planning of the fieldwork. A key element in the design-based approach to inference are the {\it design weights}, which are the reciprocals of the {\it inclusion probabilities}. We now derive these probabilities for a stratified two-stage cluster design. Suppose there are $N_h$ households in strata $h$ with $N_{hc}$ being the number of households in cluster $c$, $c=1,\dots,C_h$, $h=1,\dots,H$. Hence, $H$ is the number of strata, and there are $C_h$ clusters in strata $h$. It is decided to sample $n_h$ clusters in strata $h$ using PPS sampling, so that the first stage inclusion probability for cluster $c$ in strata $h$ is $\pi_{1hc}=n_hN_{hc}/N_h$. Consequently, every cluster has a non-zero probability of being sampled. The number of households to sample in cluster $c$ of strata $h$ is $n_{hc}$. The second stage inclusion probability for sampling is $\pi_{2hc}=n_{hc}/N_{hc}$. The overall inclusion probability is therefore,
$$\pi_{hc} = \pi_{1hc} \times \pi_{2hc}=\frac{n_hN_{hc}}{N_h} \times \frac{n_{hc}}{N_{hc}}=\frac{n_{h}n_{h,c}}{N_h}.
$$
If a constant number of clusters is sampled at the second stage, i.e.,~$n_{h,c}=n$, then we write $m_h=n_h \times n$ and obtain $\pi_{hc}=m_h/N_h$, independent of $c$. Thus inclusion probabilities are equal within strata (self-weighting), provided the size measure (number of households) is equal for each PSU in the first and second stage. In sampling with more than two stages --  such as PSUs being districts, then second stage units being enumeration areas, and third stage units being households -- the second stage units are again selected by PPS to achieve equal probabilities of selection. Multi-stage sampling tends to increase variance substantially, because most of the variance accrues at the first stage of selection. Therefore, multi-stage sampling is usually avoided if possible. For SAE, surveys that use more than two stages are not ideal, because the sample becomes concentrated in the relatively few geographic areas that have been selected in the first stage. 

In stages selected with PPS, surveys in developing countries usually employ linear systematic PPS. The method has the benefit that it is simple and allows for implicit stratification if desired. Implicit stratification involves sorting the sampling frame by some variable, usually geographic location. If the selection is systematic, it is likely that adjacent selections will be more similar to each other than to non-adjacent selections. Pairs of adjacent clusters are therefore defined as implicit strata. The DHS used implicit stratification for early surveys but do not any longer.

A great amount of energy is expended on calculating the variances of estimators.
The drawback of linear systematic PPS is that the resulting selections of PSUs are dependent on each other. The lack of independence precludes the use of some variance estimators, such as the Yates-Grundy-Sen estimator \citep{yates:grundy:53}, that require the joint inclusion probabilities of PSUs, which cannot be defined. Therefore for variance calculation purposes, it is generally assumed that PSUs were sampled with replacement, even though this is not the case under linear systematic PPS. This assumption leads to a generally insignificant inflated variance estimate. Alternative PPS selection methods do exist, see \cite{brewer:hanif:13}, but are seldom used even though statistical suites such as SPSS and SAS provide them, as do several R-packages. Surveys in developed countries often employ different designs since more information is available to the sampler.

Design based weights express the contribution of each selected sampling unit
to the estimate of a population parameter. The starting point for weight construction is the design weights. They are the inverse of the product of the inclusion probabilities at each stage of sampling. The design weights thus directly reflect the sampling process. Imperfections, such as non-response, mar most surveys. A common way to deal with non-response is to adjust the weights, by up-weighting sampling units that are assumed to be similar to the non-responding sampling unit. The adjustment is typically carried out using neighboring units, for example, a group of geographically adjacent survey clusters, as adjustment cells. The estimation weight is then the product of the design weight and the inverse of the non-response rate for the cell.  An alternative is predicting the response propensity of each sampling unit using, for example, logistic regression. The estimation weight then becomes the product of the design weight and the inverse of the predicted response probability for each sampling unit. Methods for adjusting for non-response are described in Chapter 8 of \cite{lohr:19}.

Estimates of totals from sample surveys may also differ from known totals. For example, the age and gender distribution from a survey may be different from reliable population registration.  To correct for the difference, post-stratification, raking, and calibration are sometimes used to adjust the weights so that the survey estimates match the known totals. Post-stratification divides the population into groups, for example by age and sex, and adjusts the weights within each group, so that known population totals are recovered. Raking does the same adjustment but to the marginals of the table formed by the classification variables. Calibration is similar but uses continuous variables instead of discrete.
Post-stratification, raking, and calibration ideally impart information to the sample, and therefore reduce bias.  These methods are commonly used in countries with good sources of information. They are sometimes used when the analyst believes that some sources, such as projections from a population census, are better than the survey estimates, but that is a practice that easily goes awry.

In general, weight variation in household surveys increases variance with a factor of 1+$\mbox{CV}^2$ \citep{kish:92}, where CV is the coefficient of variation of the weights.  Most household surveys strive for equal probability of selection, and therefore, constant weights at least within each stratum (as we saw with self-weighting). Equal weights are difficult to achieve in practice. Survey weight variation comes from three principal sources. The first source is disproportional allocation of sampling units to strata, which is a part of the design itself.
The second are differences that arise because of differences between estimates of the number of sampling units in one stage of the sampling process and subsequent ones. For example, in a typical two stage cluster sample where the clusters are selected with PPS, inclusion probabilities will be equal for all households in a stratum if the number of households in each cluster used for selection in the first stage is the same as the number of households found in the listing of the clusters.  Since the first stage numbers come from the sampling frame, which may be somewhat outdated, and the listing is carried out immediately before the survey, the difference between the two numbers is often large and leads to substantial weight variation. Third, weight variation may arise from non-response corrections, post-stratification, raking and calibration. 
Since the weight variation increases the variance, large weights are truncated in many surveys. The truncation trades the cost of a (hopefully) small bias for an often substantial reduction in variance, and therefore also MSE.

As a result, the final data are the product of several random processes \citep{pfeffermann:11}. The first is the generating of population units. The second is the selection resulting from the application of the sampling design. The third is the selection of responding units, given that they have been included in the sample. Finally, there is usually an ad-hoc modeling step that is used to force sample statistics to resemble population parameters and reduce variance.


\section{Traditional Methods}\label{sec:traditional}

\subsection{Direct Estimation}

A direct estimate of a quantity in a specific area and time period only uses data on the variable of interest from that area and time period. For simplicity of explanation, we assume that the strata correspond to the small areas of interest and that the weights are simply the reciprocal of the inclusion probabilities. We will also not explicitly index the clusters in this section, since we the discussion is relevant for general designs. The areas will be indexed by $i$, and let $d_{ik}$ be the design weight associated with individual $k$ in area $i$, whose response is $y_{ik}$.
 Within area $i$, the design-based weighted (direct) estimator \citep{horvitz:thompson:52,hajek:71} is
\begin{equation}\label{eq:HT}
\widehat{
m}^{\tiny{\mbox{HT}}}_i=  \frac{\sum_{k \in s_{i}} d_{ik} Y_{ik} } {\sum_{k \in s_i} d_{ik} },
\end{equation}
and its variance $V_i^\star$ may be calculated using standard methods; often the jackknife is used in a LMIC context \citep{pedersenandliu:2012}. A starting point for analysis is to map the weighted estimates. These weighted estimates have excellent properties (so long as the weights are reliable and stable), for example design consistency. If  abundant data are available, no further modeling is needed.  In general, the variance is $O(n_i^{-1})$  and so for small $n_i$ this approach will not be sufficient. In this case, models are necessary to allow for incorporation of covariate information and/or leveraging of spatial dependence between areas.

\subsection{Indirect Estimation}\label{sec:indirect}

We briefly describe a number of traditional indirect estimators, that are constructed to reduce the mean squared error (MSE), calculated under the randomization distribution. These estimators are generally design-based, but achieve a favorable MSE through variance reduction. Suppose we have covariates $\bmx_{ik}$, that are available on the complete population $k=1,\dots,N_i$, $i=1,\dots,m$. In a conventional regression analysis, attention focusses on the slope coefficients, but in SAE we are interested in $m_i$, and so it makes sense that we would want to make inferences about  the responses of unobserved individuals.
A {\it synthetic estimator} is,
$$\widehat{m}_i
^{\tiny{\mbox{SYN}}} = \frac{1}{N_i} \sum_{k=1}^{N_i} \bmx_{ik}^{\rT} \widehat \bmB,
$$
where,
$$\widehat \bmB= \left[ \sum_{i=1}^m \sum_{k \in S_i}d_{ik} \bmx_{ik}^{\rT}\bmx_{ik} \right]^{-1} \sum_{i=1}^m \sum_{k \in S_i} d_{ik}\bmx_{ik}^{\rT}y_{ik},$$
are the slope estimates derived from the set of samples across {\it all} areas.
The success of this estimator depends on how appropriate the regression model is for all areas. 
The variance will be $O(1/n)$, where $n=\sum_{i=1}^m n_i$ is the total sample size, but the possibility of large bias leads to this estimator being unpopular.
 Fitting separate models in each area is appealing, but then the small $n_i$ problem re-emerges. As a side note, one possible compromise would be to use spatially varying coefficient models in which each area has its own slope that is smoothed across areas.

To deal with the potential large bias, the bias may be estimated to give the {\it survey-regression estimator}:
\begin{eqnarray*}
\widehat{m}_i
^{\tiny{\mbox{SR}}}&=& \frac{1}{N_i} \sum_{k=1}^{N_i} \bx_{ik}^{\rT} \widehat \bmB + \frac{1}{N_i}
\sum_{k \in S_i} w_{ik} (y_{ik} - \bx_{ik}^{\rT}  \widehat \bmB)
\\
&=&\widehat{m}^{\tiny{\mbox{HT}}}_i + (\overline{\bx}_i - \widehat{\overline{\bx}}^{\tiny{\mbox{HT}}}_i)^{\rT} \widehat \bmB
\end{eqnarray*}
where $\widehat{m}^{\tiny{\mbox{HT}}}_i$ and $\widehat{\overline{\bx}}^{\tiny{\mbox{HT}}}_i$ are the Horvitz-Thompson estimates of $m_i$ and $\overline{\bx}_i$. This estimator is approximately design-unbiased but 
the variance is unfortunately back to $O(1/n_i)$.
A composite estimator is of the form
$$\widehat{m}_i^{\tiny{\mbox{COM}}} = \delta_i \widehat{m}_i^{\tiny{\mbox{SR}}}  + (1-\delta_i) \widehat{m}_i^{\tiny{\mbox{SYN}}}
$$
with $0 \leq \delta_i \leq 1$ estimated in such a way that for larger $n_i$ we have  $\delta_i$ closer to 1. This estimator is intuitively appealing, and when motivated by a random effects model leads to a principled approach to estimation of $\delta_i$.

\section{Spatial Smoothing Models}\label{sec:spatial}

\subsection{Area-Level Models}\label{sec:arealevel}

In this section and the next we focus on binary outcomes, because the Malawi HIV prevalence example falls under this umbrella.
In a major advance,  \cite{fay:herriot:79} introduced a very clever approach that models a transform of the weighted estimate, in order to gain precision by using a random effects  model. For binary outcomes, one choice of transform is
$Z_i= \mbox{logit} \left(\widehat m_i^{\tiny{\mbox{HT}}}  \right)$,
and let the associated design-based variance estimate be denoted $V_i$.  An area-level model is,
\begin{eqnarray}
Z_i &\sim & \mbox{N}(\theta_i , V_i)\label{eq:smoothed1}\\
\theta_i &=&  \bx_i^{\tiny{\mbox{T}}} \bbeta + e_i + S_i,\label{eq:smoothed2}
\end{eqnarray}
where  $\theta_i$ is the  logit of the true proportion in area $i$. Area-specific deviations from the regression model are modeled using a pair of random effects. The independent and identically distributed (iid) terms are  $e_i \sim_{iid} \mbox{N}(0,\sigma_\epsilon^2)$ while  the collection $\bmS=[S_1,\dots,S_n]$ are assigned a spatial distribution. The Fay-Herriot model  did not include the spatial random effects $\bmS$, but the iid random effects only. Choices for the spatial distribution
are described in \cite{banerjee:etal:15}, with common forms being the conditional autoregressive (CAR) and intrinsic CAR (ICAR) models. Both of these choices captures the concept that, in general,  many outcomes are likely to be similar in locations that are close-by.  \cite{mercer:etal:15} and \cite{li:etal:19} used these {\it smoothed direct} models in a space-time context, using spatial ICAR  and temporal random walk components, along with a space-time interaction term. Both the CAR and ICAR choices are Markov random field (MRF) models, which offer computational advantages \citep{rue:knorrheld:05}. 
This model may be fit using a frequentist approach or a Bayesian approach in which  priors are placed on the fixed effects (the $\bmbeta$'s) and on the variances/spatial dependence parameters. Design-based consistency is achieved (so long as the priors don't assign zero mass to the true $\theta_i$) since the $V_i$ term will tend to zero and the bias due to the random effects smoothing disappears asymptotically.

Two practical difficulties with this approach are that the direct estimates may be on the boundary for a summary parameter that is not on the whole real line. For example, in the binary case we may have $\widehat m_i^{\tiny{\mbox{HT}}}$ equal to 0 or 1. In this case, $Z_i$ will be undefined. Further, a transform of the weighted estimator may not share the same design-based properties as the untransformed estimator, such as being design unbiased. These problems may be alleviated by using an unmatched sampling and linking model \citep{you:rao:02}. 
A second difficulty  is that reliable variance estimates $V_i$ may be unavailable, particularly for areas with few/no samples. In this case, variance smoothing models can be used \citep[Section 6.4.1]{rao:molina:15}. 


\subsection{Unit-Level Models}\label{sec:unitlevel}

In this section, we assume the units of analysis are clusters within a multi-stage cluster design. We  let $\bms_{ic}$ represent the geographical location of cluster $c$ in area $i$, and explicitly index the counts and sample sizes as $Y_{ic}$ and $n_{ic}$, respectively. A crucial assumption here (Section 4.3 of \cite{rao:molina:15}) is that the probability of selection, given covariates, does not depend on the values of the response (as discussed in the context of {\it ignorability} in Section \ref{sec:notation}). This implies that if stratified 
random sampling is used, stratification variables must be included in the model.
One would expect cluster sampling to lead to correlated responses within clusters, and cluster-level random effects are introduced to accommodate this aspect \citep{scott:smith:69}. Another situation in which care is required is when PPS sampling is carried out. If the model used mis-specifies the relationship between the response and size then incorrect inference will result. To provide some protection against this, \cite{zheng:little:03} and \cite{chen:etal:10} model the response as a spline function of size for continuous and binary outcomes, respectively.

For a binary response, a common model \citep{diggle:giorgi:19} is,
\begin{eqnarray*}
Y_{ic} | p_{ic} &\sim& \mbox{Binomial}(n_{ic},p_{ic}).
\end{eqnarray*}
Care must be taken when modeling  $p_{ic}=p(\bms_{ic})$, the response probability associated with location $\bms_{ic}$. For a stratified design it is, in general, important to include fixed effects for each strata \citep{paige:etal:20}. In practice, there are a number of options when we have a design with strata that consist of urban/rural crossed with geographical districts. If there were no random effects, then strictly we would need  interactions between district and urban/rural. However, the models we describe include spatial random effects (to leverage spatial dependence) and interactions would produce an unidentifiable model. Consequently, we include a single urban/rural effect and so are tacitly assuming that the association between the response and the urban/rural variable is approximately constant across areas.  Another possibility, would be to include a spatially varying urban/rural coefficient, or model the associations separately in larger regions, in the spirit of what is sometimes done with synthetic estimation.  One candidate model is,
\begin{eqnarray}
p_{ic}  &=&\mbox{expit} \left(  \beta_0+ \bx(\bms_{ic})^{\tiny{\mbox{T}}} \bbeta_1 +  z(\bms_{ic}) \gamma+S(\bms_{ic}) + \epsilon_{ic} \right)
\end{eqnarray}
where  $z(\bms_{ic})$ is the strata within which cluster $c$ lies, $\exp(\gamma)$ is the associated odds ratio, $\bx(\bms_{ic})$ are covariates available at location $\bms_{ic}$, with odds ratios $\exp(\bbeta_1)$. The spatial random effect $S(\bms_{ic})$ is associated with cluster location $\bms_{ic}$, and may be continuous or discrete. The cluster-level error $\epsilon_{ic} \sim \mbox{N}(0,\sigma_\epsilon^2)$ is the so-called {\it nugget}, which is traditionally taken to represent short scale variation and/or ``measurement error''.
A model-based geostatistical model takes $S(\bms_{ic})$, as a realization of a zero-mean Gaussian process (GP). GP models are common choices for continuous spatial models and imply that any collection of spatial random effects have a multivariate normal distribution. A popular choice for the variance-covariance \citep{stein:99} is the Mat\'ern covariance function, for which the covariance between is,
$$
	\mbox{cov}(S(\bms_1),S(\bms_2)) = \sigma_\mathrm{S}^2 \frac{2^{1-\nu_\mathrm{S}}}{\Gamma(\nu_\mathrm{S})} \left(\sqrt{8\nu_\mathrm{S}}\frac{||\boldsymbol{s}_2-\boldsymbol{s}_1||}{\rho_\mathrm{S}}\right) \mathrm{K}_{\nu_\mathrm{S}}\left(\sqrt{8\nu_\mathrm{S}}\frac{||\boldsymbol{s}_2-\boldsymbol{s}_1||}{\rho_\mathrm{S}}\right),
$$
where $\rho_\mathrm{S}$ is the spatial range corresponding to the distance at which the correlation is approximately 0.1, $\sigma_\mathrm{S}$ is the marginal standard deviation,
$\nu_\mathrm{S}$ is the smoothness (which is usually fixed, since it is difficult to estimate), and $\mathrm{K}_{\nu_\mathrm{S}}$ is a modified Bessel function of the second kind, of order $\nu_{\mathrm{S}}$. When the number of clusters $C$ is large, computation is an issue, because we need to manipulate $C \times C$ matrices which involves $O(C^3)$ operations \citep{rue:knorrheld:05}. Various approximations have been proposed to overcome this problem, for example, the stochastic partial differential equations (SPDE) approach pioneered by \cite{lindgren:etal:11} -- this is the approach we use in Section \ref{sec:HIV}. Other approaches are described by \cite{heaton:etal:18}. 
 
Aggregation to the area-level is carried out via,
\begin{equation}\label{eq:area} p_i = \int_{A_i} p(\bms) \times q(\bms)~d\bms \approx \sum_{l=1}^{M_i} p(\bms_l) \times q(\bms_l)
\end{equation}
where $p(\bms) = \mbox{expit}( \beta_0+ \bmx(\bms)^{\tiny{\mbox{T}}} \bbeta_1+z(\bms_{ic}) \gamma + S(\bms))$ is the risk at location $\bms$ (the nugget is, for better or worse, frequently left out, since it is viewed as measurement error) and  $q(\bms)$ is  the population density at $\bms$, which is needed at all locations on the approximating mesh, $\bms_l$, $l=1,\dots,M_i$. Cluster level covariate models are appealing when compared to area-level covariate alternatives, since they are closer to the mechanism of action, and reduce the possibility of ecological bias \citep{wakefield:08}. A large disadvantage in an SAE context is that to aggregate to the area-level, at which inference is required, one needs to know the values of the covariates on the mesh, including the design variables. Obtaining reliable population density and covariate surfaces is not straightforward.
The direct and smoothed direct approaches avoid the need for a population surface, since the weights implicitly include population information from the original sampling frame.

An alternative, overdispersed binomial, unit-level model that we use for the HIV prevalance data is,
\begin{eqnarray}
Y_{ic} | p_{ic},\lambda &\sim& \mbox{BetaBinomial}(n_{ic},p_{ic},\lambda )\label{eq:BB1}\\
p_{ic} &=&\mbox{expit} \left(  \beta_0+ \bx_{i}^{\tiny{\mbox{T}}} \bbeta_1 +  z_{ic} \gamma+ e_i  + S_i \right)\label{eq:BB2}
\end{eqnarray}
where $\lambda$ is the overdispersion parameter and we have taken the spatial random effect to be decomposed as $S(\bms_{ic})=e_i + S_i$, with $e_i$ and $S_i$ as in (\ref{eq:smoothed2}). The marginal variance is
$\v(Y_{ic}) = n_{ic}p_{ic}(1-p_{ic})[1+(n_{ic}-1)\lambda]$ so that small values of $\lambda$ correspond to little overdisperison.
The aggregate risk is far easier to evaluate than the continuous spatial model, i.e.,~as calculated via (\ref{eq:area}), since all clusters in the same strata have identical risk, to give,
\begin{equation}
p_i = (1-q_i)\times  \mbox{expit}( \beta_0+ e_i  + S_i ) +q_i\times  \mbox{expit} ( \beta_0+ \bx_{i}^{\tiny{\mbox{T}}} \bbeta_1+ e_i  + S_i ) \label{eq:aggmesh}
\end{equation}
where $q_i$ is the proportion of the relevant population in the area that is urban.
Dyed-in-the-wool ,design-based aficionados are wary of models such as the ones described in this section, because establishing design consistency is very difficult.


\section{SAE for HIV Prevalence Mapping}\label{sec:HIV}

We return to the HIV prevalence example, and fit both area-level and unit-level models. All fitting was done in the {\tt R} programming environment.  The smoothing models were fit using the {\tt INLA} package, which provides Bayesian inference via a fast implementation using the method described in \cite{rue:etal:09} and \cite{rue:etal:17}. To obtain the weighted estimates and variances for the areas, the {\tt survey} package \citep{lumley:04} was used. A number of the models can be conveniently fitted in the {\tt SUMMER} package \citep{martin:etal:18}. For the discrete spatial model we use the BYM2 parameterization \citep{riebler:etal:16} in which we have $b_i=e_i+S_i$ (in equation (\ref{eq:smoothed2})) and an overall variance parameter $\sigma^2_b$ for $b_i$, and a parameter, $\phi$, that represents the proportion of the variance that is spatial. In all analyses we use penalized complexity (PC) priors \citep{simpson:etal:17}, details of which may be found in the supplementary materials.

\subsection{Area-Level Models}

In high-income countries, the census (and other sources) provide variables on all of the population in small areas of interest. In this case, weighted estimates can be obtained from model-assisted procedures, such as those described in Section \ref{sec:indirect}, which can be smoothed, if needed. Unfortunately, in LMIC, there are few candidate variables, and so instead we use the smoothed direct model (\ref{eq:smoothed1})--(\ref{eq:smoothed2}) and include as a covariate the (logit of) the prevalence estimates obtained from antenatal care (ANC) facilities. Specifically, we use data from 2016, taken from Malawi Ministry of Health quarterly integrated HIV program reports. This covariate corresponds to the HIV prevalence of pregnant women attending health facilities to receive ANC, aggregated to the district level (a map of this covariate can be found in the supplementary materials). The left panel of Figure \ref{fig:various} shows the logits of the direct estimates against the logits of the ANC prevalence estimates, with a reference line added. Not surprisingly, we see a strong association.

%
%

The smoothed direct model estimates without the ANC covariate are shown in the top middle panel of Figure \ref{fig:Malawi1}. The shrinkage (overall via the iid random effects, and locally via the ICAR random effects) is apparent, with a flatter map compared to the direct estimates. The posterior uncertainty is reduced (see supplementary materials). The average standard error of the direct estimates is 0.018 and the average posterior standard deviations under the smoothed direct model is reduced by 17\%. When moving from the no covariate, smoothed direct model to the model including the ANC covariate, the posterior median of $\sigma_b$ (reflecting the amount of residual variation) changes from 0.40 to 0.14, and $\phi$ (the proportion of spatial variation) goes from 0.56 to 0.26. The odds ratio (95\% credible interval) associated with the logit ANC covariate is 2.8 (2.0, 3.8) so the association is very strong, which explains why the across-district residual variability is so reduced when the covariate is added. The strong spatial structure of the covariate explains why $\phi$ is reduced. The dream of spatial modeling is to find covariates that make the area-level random effects ``go away". The top right panel of Figure \ref{fig:Malawi1} shows the estimates from the smoothed direct model that includes the ANC covariate.
The supplementary materials contain comparison plots of the estimates and their uncertainties under various models. The smoothed direct model with the ANC covariate has a 39\% reduction in the uncertainty, when compared to the direct estimates.
The pattern of lower prevalence in the north, with higher levels in the south is still apparent in the smoothed estimates.
Even with the spatial smoothing and use of the covariate, the uncertainty in the ranking of areas is still large. The supplementary materials contains distributions on the rankings under  six different models. Mulanje consistently comes out as the district with the highest HIV prevalence among women ages 15--29.

\begin{figure}[htbp]
    \centering
        \includegraphics[width=0.4\linewidth]{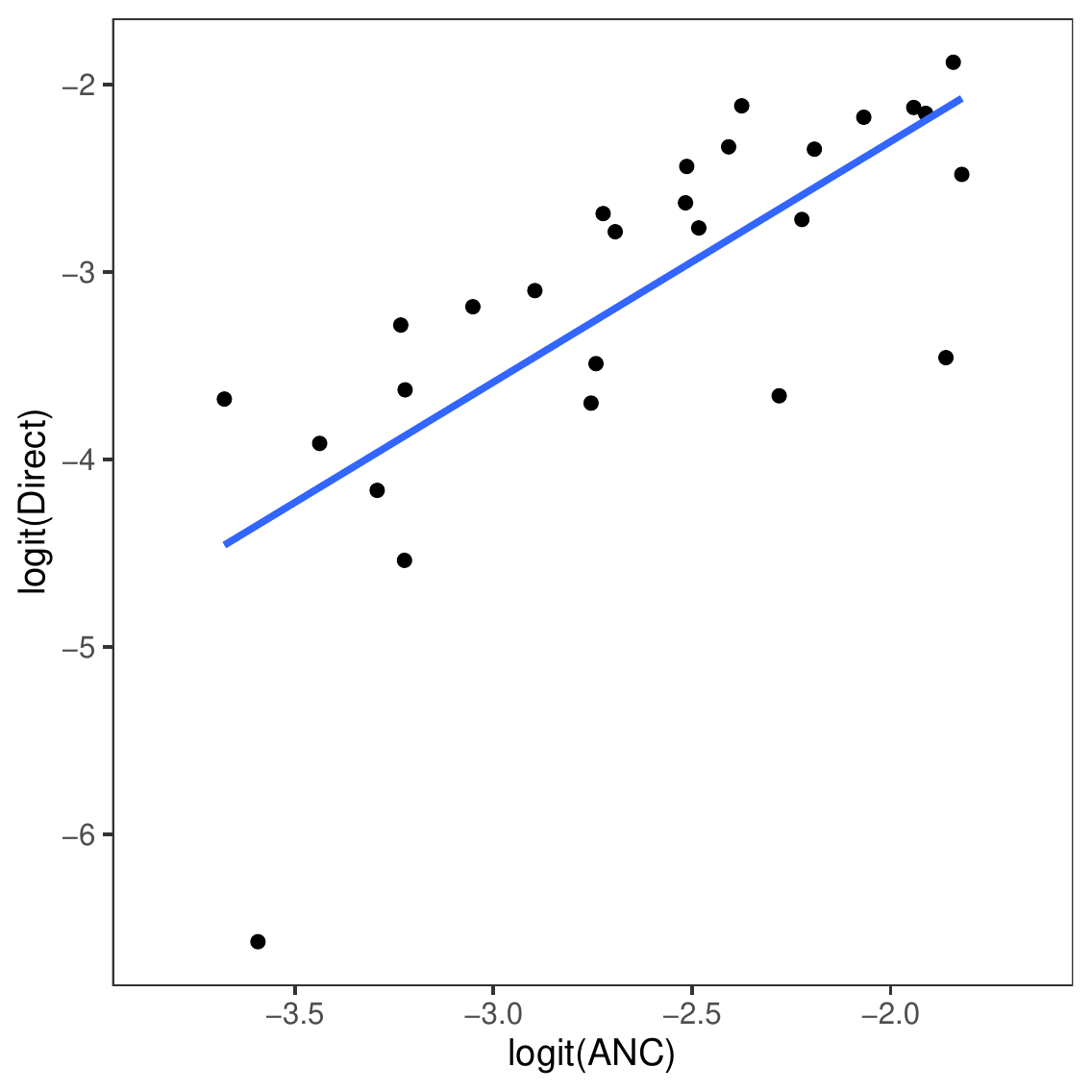}
    \includegraphics[width=0.45\linewidth]{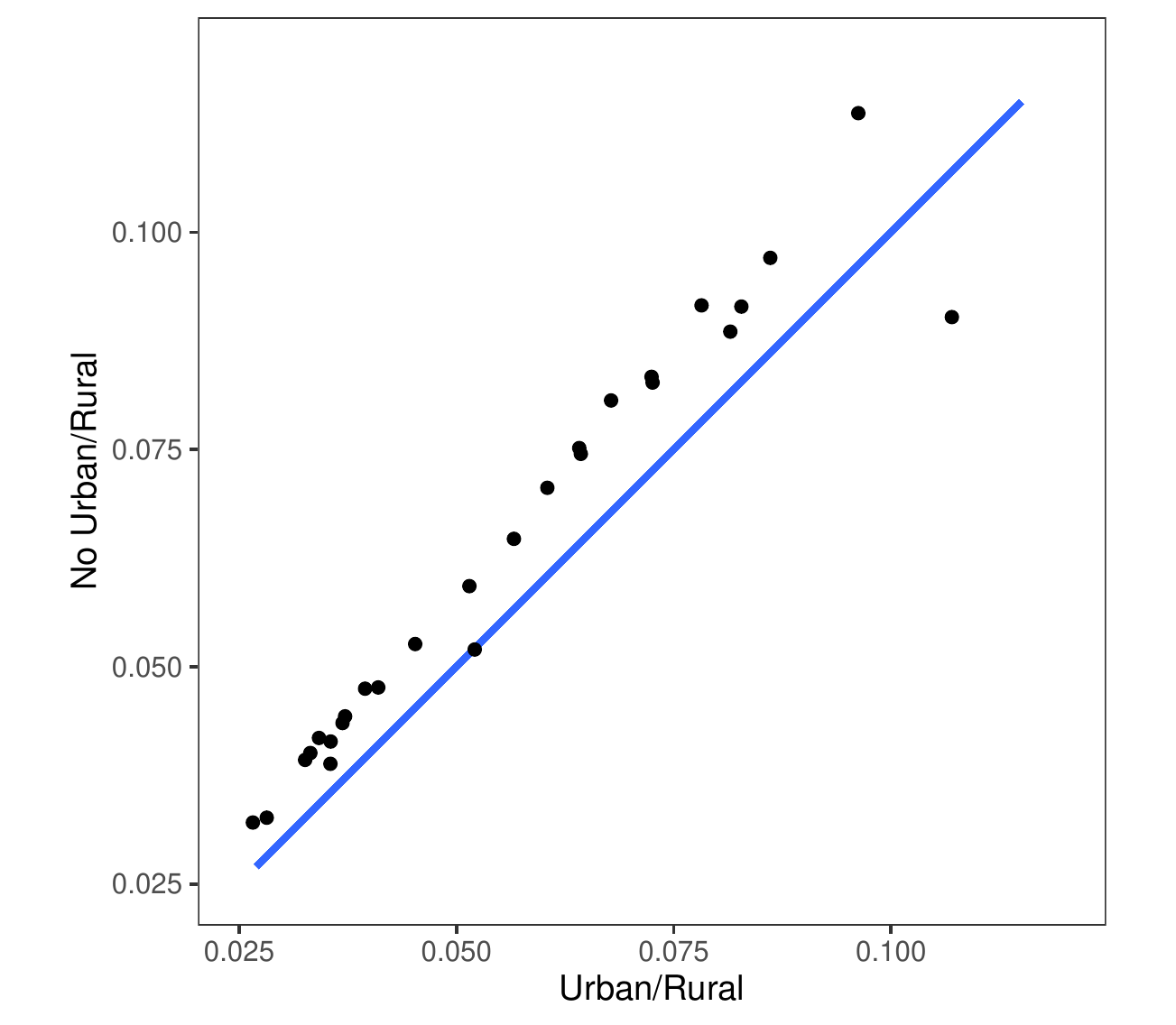}
    \caption{Left: logit of direct prevalence estimates versus logit of ANC prevalence estimates -- we see a strong association. Right: District prevalence estimates from two unit-level models. On the y-axis, the prevalence estimates are from a model with no urban/rural adjustment, while on the x-axis the model has an adjustment. The estimates from the no adjustment model are too high because of the oversampling of urban areas, which have higher HIV prevalence.}
    \label{fig:various}
\end{figure}

\subsection{Unit-Level Models}

%
%
We fit three betabinomial models (\ref{eq:BB1})--(\ref{eq:BB2}): (1) no urban/rural and no covariates, (2) urban/rural and no covariates, (3) urban/rural and ANC covariate. Model (1) should not be used, as it does not account for the urban/rural stratification; we include to demonstrate the bias this introduces. The bottom row of Figure \ref{fig:Malawi1} shows the fits from these three models. Again, the smoothness compared to the direct estimates is apparent, and the fits are quite similar to the smoothed direct models. Closer examination reveals differences, however. The supplementary materials gives comparisons of all the estimates. It is clear that leaving the urban/rural covariate out of the model leads to bias. From the sampling frame we know the number of clusters that are urban and rural (this information is given in the supplementary materials) which can be directly used for the aggregation to the admin-2 level, via (\ref{eq:aggmesh}).

 In the no covariate model the odds ratio associated with being urban is 2.3 (1.8,2.9) and this hardly changes when the ANC variable is added. This more than doubling of the odds  is a cluster level association. The odds ratio associated with the ANC covariate is 2.3 (1.7, 3.1), so the association is also strong. The reductions in the total spatial residual terms, and the proportion of this variance that is spatial both plummeted when the ANC covariate was included in the model.
As compared to the average standard errors of the direct estimates, the average posterior standard deviations for models (1)--(3) above were 18\%, 29\% and 43\% lower., respectively So there is a greater reduction in uncertainty under the unit-level models than under the area-level models. The estimates from model (1) are clearly biased (they are too high) because of the non-adjustment for urban/rural. For the three models the posterior medians of $\sigma_b$ (representing the standard deviation  of the residual variation on the logistic scale) are 0.39, 0.38, 0.18. The proportion of the variation that is spatial is 0.60, 0.67, 0.30. Hence, we see the same pattern as the area-level modeling with the inclusion of the ANC covariate greatly reducing the residual uncertainty and the proportion that is spatial.


We also calculate estimates for the 243 admin-3 areas, which is more difficult  with the smoothed direct model, because of data sparsity (26 of the areas contain no data). The supplementary materials contain summaries of this analysis. These results should be viewed with some caution, since model-checking at this level is very difficult, and the aggregation depends on knowing the proportion of each admin-3 area that is urban. Finally, we fit the continuous spatial model, with and without the area-level covariate and produce pixel-level estimates of the HIV prevalence. This analysis should also be viewed with caution, particularly for the admin-2 summaries, since the aggregation required a mesh specific estimate of urban/rural. We use WorldPop population density estimates \citep{stevens:etal:15,wardrop:etal:18}. Specifically, using the known proportion of clusters that are labeled as urban in the sampling frame we create a population density threshold (to give the correct urban proportion of the population when a geographic partition is formed, based on this threshold), and then we can obtain an urban/rural mesh to use in (\ref{eq:aggmesh}). Results of these analyses are contained in the supplementary materials.

There are a number of outstanding issues with the unit-level model-based analyses that need further consideration. In general, the weights have three components: design, non-response and post-stratification/raking and one needs to consider all three. We have discussed the stratification aspect, but further thought is required for non-response, while post-stratification from a modeling perspective has been considered 
 \citep{gelman:07}. In practice, the urban/rural covariate is not known with any great accuracy geographically, and the impact of this on inference needs to be investigated.

\subsection{Model Assessment}

One of the hardest parts of SAE is model assessment. The direct estimates, $\widehat{m}^{\tiny{\mbox{HT}}}_i$  if available, provide one standard for 
comparison. Asymptotically, $\widehat{m}^{\tiny{\mbox{HT}}}_i \sim \N( m_i,V^\star_i)$. One strategy is to systematically remove one area at a time, and then obtain a prediction of the missing area's prevalence, based on the remaining 26 districts. The supplementary materials contain plots of the direct estimates versus the predicted estimates, and we see systematic problems with the models that leave out the urban/rural covariate. The performance of both the smoothed direct (area-level) model with the ANC covariate and the unit-level model with urban/rural and ANC prevalence is superior to the other models.
The DIC and log CPO summary measures supported this conclusion. Checking the admin-3 estimates is far more difficult because there is no reliable direct estimates for comparison in a cross-validation exercise.

\section{Application of SAE Techniques to COVID-19 Modeling}\label{sec:COVID}

Coronavirus disease (COVID-19) is an infectious disease caused by the SARS-CoV-2 virus. There are many endpoints that are of interest in geographical studies including: disease prevalence, disease incidence, the number and fraction of people who have been infected with SARS-CoV-2 (via an antibody test), measures of social distancing, and excess mortality. Estimating each of these responses over time and by age, gender, and race is also highly desirable. Many study/data types would not fall under the heading of SAE because they are based on routine surveillance data and are nominally a complete enumeration. Surveillance data are often examined via space-time {\it disease mapping} or {\it cluster detection} techniques \citep{waller:gotway:04}.

Seroprevalence studies, in which blood samples are taken from sampled individuals and examined for antibodies, have been carried out to estimate the number of infections due to SARS-CoV-2. With an appropriate sampling design, and a reliable test, such studies can be used to estimate the proportion of the population, by demographic groups, who have been previously infected by SARS-CoV-2. These studies can be used for SAE, if the numbers sampled are sufficiently large. \cite{stringhini:etal:20} describe a seroprevalence study carried out in Geneva, Switzerland, with participants taken from an existing study that contained a stratified sample from the study population (the canton of Geneva). The selection of individuals is key since, for example, those reporting at health facilities (who are subsequently tested) are not representative of the population.  Unfortunately, many of the early seroprevalence studies came (justifiably) under fire for data quality issues and for sampling from ill-defined populations, which leads to great difficulties with interpretation. This is well-documented by \cite{vogel:20}, who gives a number of examples. Health researchers are under intense pressure, given they are working in the context of the political and press feeding-frenzy for results to be made available. This has lead to inadequate validation of the antibody tests used and limited scientific scrutiny of the design of studies through peer-review. 



It has been documented that certain subgroups (for example, based on age, sex, race) are disproportionately affected by COVID 19, and as such, a stratified analysis may be used to reliably estimate outcomes of interest.
But the numbers required will be large if a small area analysis is envisaged, given the relative rarity of death from COVID-19, in particular.

A potential drawback with traditional survey sampling and SAE in the context of COVID-19 is the possibility that seroprevalence or incidence is highly clustered. There are two notions of clustering that are important here. The first is infectious network clustering, i.e.,~that an infected person infects others in his or her network. Such clustering, while intrinsic to any infectious disease, may or may not lead to the other form of clustering, namely geographic clustering. Geographic clustering of COVID-19 is common, both in communities and in institutions (there are many factors which may lead to this, including the propensity for individuals of common demographic groups, or having high-risk, tending to live in close proximity).  When geographic clustering is present, a large proportion of the infected population may be located in relatively few survey clusters, many of which may be geographically adjacent.  Since a typical cluster sample design spreads out the sample clusters over a geographic region, there is a large chance that the sample will not pick up the high prevalence areas. One may solve the problem by increasing the sample size, but that rapidly becomes costly. 

An alternative is adaptive cluster sampling \citep{thompson:92, thompson:seber:96}.  An ordinary cluster sample is first drawn. Then, each selected cluster is listed and target individuals identified (e.g.,~persons with a positive test for Sars-CoV-2). If the number of target individuals is larger than a given threshold, then clusters that share a border with the initial cluster are included in the survey. The process is repeated for each of the adjacent clusters, until a network of clusters is formed that has no adjacent cluster thst satisfies the threshold. 

If the threshold is well selected, then the adaptive cluster sample will pick up geographically based networks of infection well. If it is set too low, then all clusters in the frame will be included. If the threshold is too high, the sample limits itself to the initial sample.  With a target population that is clustered and judicious setting of the threshold value, adaptive cluster sampling can be very efficient. Unfortunately, there is no optimal sampling strategy for adaptive cluster sampling \citep{turk:borkowski:05,thompson:seber:96}, so the exact choice of design must be piloted.

A potential disadvantage of adaptive cluster sampling is that while it may be efficient for the estimator that informs the design, such as the prevalence of a disease, it is not necessarily efficient for estimators of other population quantities. Thus, if the survey aims to estimate additional quantities other than the one designed for, one might have to consider other designs. This problem is, to some extent, mitigated by the fact that many other quantities may be both easy to estimate and not require narrow confidence bands. For example, the proportion of people using face masks may not require as precise an estimate as a seroprevalence estimate.

Looking into the future, given the successful development of a vaccine, one could use SAE techniques to estimate vaccination coverage rates. Such an approach has been used in many different settings: BCG, DPT, OPV, and measles in India
\citep{pramanik:etal:15} and DPT  in Africa \citep{mosser:etal:19}.
In the United States (US), in the absence of a national register, particular outcomes can be geographically examined using survey data.
\cite{albright:etal:19} carried out SAE  at the county level in Alabama for human papillomavirus (HPV) in counties of Alabama using data from the National Immunization Survey.  
 
To get at transmission dynamics, one needs to fit more complex models such as susceptible-exposed-infected-removed (SEIR) models. Unfortunately, there are a number of serious drawbacks for using such models in a SAE enterprise. The first is that the  data required are not routinely available, with the usual problems of under- and mis-reporting being a serious problem. The second is that fitting stochastic models is very challenging, see the references in \cite{fintzi:etal:17}. To avoid the computational difficulties one may use deterministic compartmental models \citep{anderson:may:91} and then map summary statistics such as the reproductive number $R_0$. In large populations such models may perform well, but  such models may perform poorly when the system is far from its deterministic limit \citep{andersson:britton:00}; inference is also difficult because error terms are ``added on" rather than the stochastic and deterministic parts of the models being entwined, as is the ideal for a statistical model. However, such models are far preferable to simple curve-fitting exercises in which some generic  function is fit to data -- such approaches are likely to perform very poorly for predictions of complex phenomena because they (basically) contain little of the science or relevant context. The impact of interventions, such as social distancing and vaccination campaigns is also impossible to determine with such approaches.


%
%

\section{Discussion}\label{sec:discussion}

SAE is a huge topic and we have been selective in this review, focusing on the use of Bayesian spatial models for analyzing survey data. 
There is a disconnect between the spatial statistics and SAE research communities. The former are very cavalier about ignoring the survey design and the latter tend to use simple discrete spatial models (or include iid spatial random effects only) and empirical Bayes estimation. Regular spatial modeling with binary data does not emphasize consistency, perhaps because the parameters of nonlinear random effects models are not generally consistent, unless the model is correctly specified. Outside of the linear model, results on frequentist (model-based) consistency for the parameters of spatial models are few and far between.

It is desirable that when area-level estimates are aggregated they are in agreement with estimates for larger areas within which they are nested. The estimates in the larger areas may be well estimated due to larger sample sizes and additional data. Such benchmarking is an important endeavor in many applications. For example, \cite{li:etal:19} produced small area estimates of under-5 mortality in 25 countries using DHS data and benchmarked to the official UN national estimates, which use far more extensive data sources. Many approaches have been suggested. In particular, we find the Bayesian method of \cite{zhang:bryant:20} very appealing, and it could be used with the area-level and unit-level models we have described.

We have discussed both discrete and continuous spatial models. The former have dominated the SAE literature, and have many appealing features including computational efficiency, a wealth of experience in their use, and robustness to different data generating mechanisms \citep{paige:etal:20}.  However, the discrete spatial models always have an ad hoc neighborhood specification, which is unfortunate. Continuous spatial model are far more appealing in this respect, and also allow data that are aggregated to different levels to be combined \citep{wilson:wakefield:20}. But continuous spatial models have greater computational challenges, and careful prior specification is required \citep{fuglstad:etal:19}.




There is a growing literature on accounting for errors in the covariates used for modeling in the SAE literature with various  errors-in-variables models being considered \citep{ybarra:lohr:08,barber:etal:16,burgard:etal:19}; see Section 6.4.4 of Rao and Molina (2015)\nocite{rao:molina:15} for a review. 

Finally, as more and more data streams become available, there is a growing need for methods that combine data in appropriate ways, and this is especially true of SAE. For example, in a LMIC context, data may be available from a variety of surveys with different designs, sample vital registration systems and censuses. Synthesizing such data is not straightforward but can offer great benefits if carried out successfully.


\vspace{.3in}
\noindent
{\bf Acknowledgments:} The authors would like to thank the Malawi Ministry of Health for allowing the use of the ANC data, and Jeff Eaton for helpful comments on the manuscript.

\bibliographystyle{natbib} 
\bibliography{SAE-Review}
\clearpage

\section*{Supplementary Materials}

\subsection*{Malawi DHS Details}
\begin{figure}[htbp]
	\centering
	\includegraphics[width=0.8\linewidth]{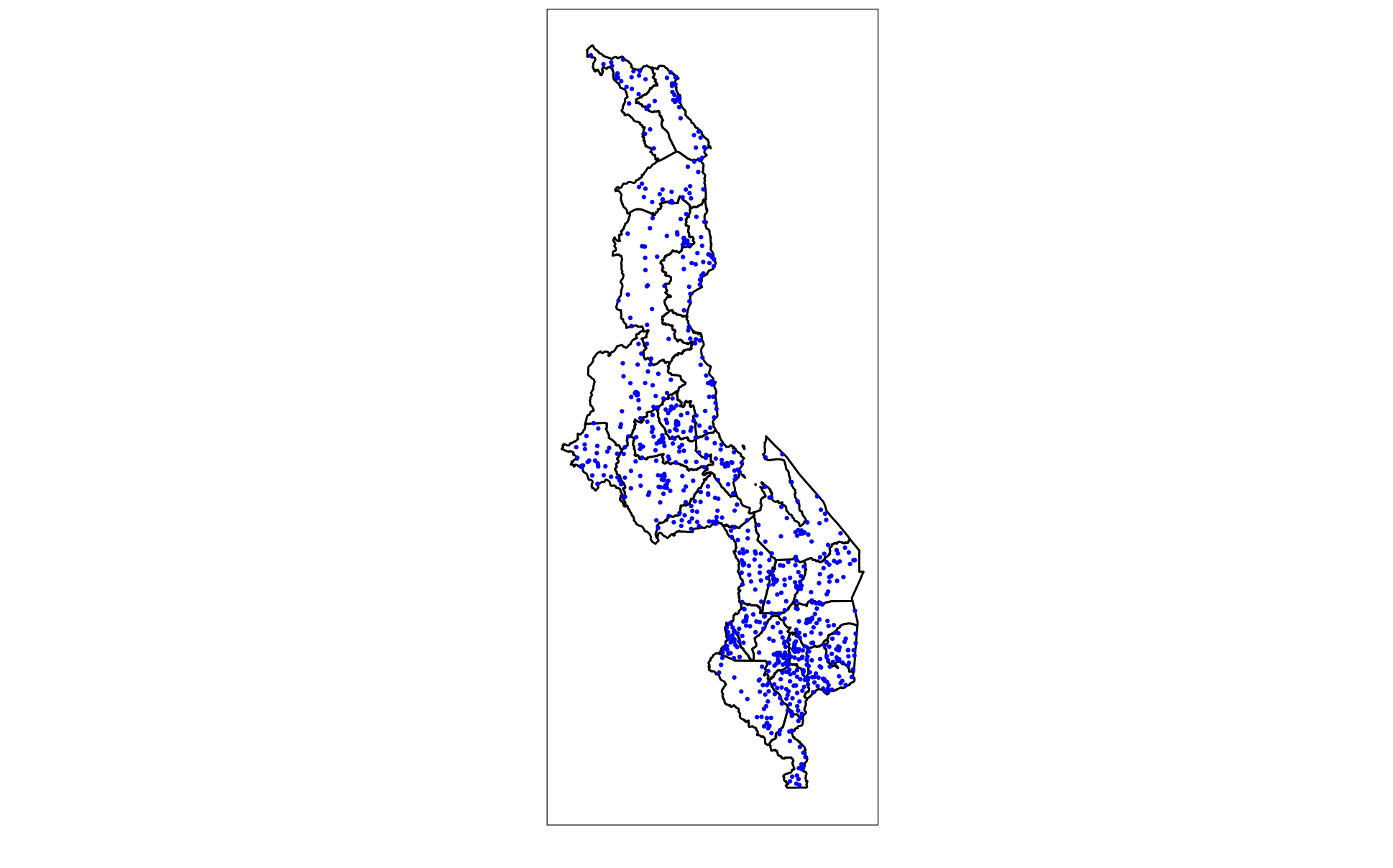}
	\caption{Locations of sampled  clusters in the Malawi 2015--16 DHS.}
	\label{fig:cluster_location_map}
\end{figure}

{ \renewcommand{\arraystretch}{0.5}
	\begin{table}[!ht]
		\centering
		\begin{tabular}{l|rr|rr|rr}
			\hline
			\multicolumn{1}{l|}{\multirow{2}{*}{Region}} & \multicolumn{1}{c}{\multirow{2}{*}{HIV Positive}} & \multicolumn{1}{c|}{\multirow{2}{*}{No. Tested}} & \multicolumn{2}{c|}{Sampled Clusters}                                                                  & \multicolumn{2}{c}{Sampling Frame}                                                            \\
			\multicolumn{1}{c|}{}                        & \multicolumn{1}{c}{}                                     & \multicolumn{1}{c|}{}                               & \multicolumn{1}{c}{Urban} & \multicolumn{1}{c|}{Rural} & \multicolumn{1}{c}{Urban} & \multicolumn{1}{c}{Rural} \\ \hline
			Balaka                                      & 13                                                       & 176                                                & 6                                            & 24                                           & 17                                            & 275                                           \\
			Blantyre                                    & 19                                                       & 185                                                & 19                                           & 16                                           & 412                                           & 381                                           \\
			Chikwawa                                    & 4                                                        & 136                                                & 4                                            & 27                                           & 16                                            & 380                                           \\
			Chiradzulu                                  & 10                                                       & 132                                                & 2                                            & 27                                           & 2                                             & 334                                           \\
			Chitipa                                     & 3                                                        & 109                                                & 5                                            & 20                                           & 11                                            & 205                                           \\
			Dedza                                       & 5                                                        & 182                                                & 5                                            & 29                                           & 15                                            & 486                                           \\
			Dowa                                        & 6                                                        & 190                                                & 5                                            & 28                                           & 18                                            & 450                                           \\
			Karonga                                     & 9                                                        & 143                                                & 8                                            & 20                                           & 37                                            & 370                                           \\
			Kasungu                                     & 7                                                        & 180                                                & 7                                            & 26                                           & 29                                            & 486                                           \\
			Lilongwe                                    & 15                                                       & 220                                                & 14                                           & 23                                           & 458                                           & 1173                                          \\
			Machinga                                    & 15                                                       & 171                                                & 5                                            & 27                                           & 19                                            & 436                                           \\
			Mangochi                                    & 22                                                       & 206                                                & 6                                            & 29                                           & 25                                            & 614                                           \\
			Mchinji                                     & 4                                                        & 184                                                & 5                                            & 26                                           & 12                                            & 374                                           \\
			Mulanje                                     & 23                                                       & 165                                                & 4                                            & 29                                           & 17                                            & 658                                           \\
			Mwanza                                      & 12                                                       & 122                                                & 6                                            & 19                                           & 9                                             & 80                                            \\
			Mzimba                                      & 6                                                        & 204                                                & 11                                           & 24                                           & 122                                           & 825                                           \\
			Neno                                        & 12                                                       & 153                                                & 3                                            & 23                                           & 3                                             & 157                                           \\
			Nkhata Bay                                  & 6                                                        & 139                                                & 5                                            & 22                                           & 12                                            & 229                                           \\
			Nkhotakota                                  & 9                                                        & 163                                                & 6                                            & 22                                           & 16                                            & 177                                           \\
			Nsanje                                      & 9                                                        & 124                                                & 6                                            & 21                                           & 14                                            & 241                                           \\
			Ntcheu                                      & 11                                                       & 164                                                & 4                                            & 28                                           & 11                                            & 468                                           \\
			Ntchisi                                     & 1                                                        & 145                                                & 4                                            & 23                                           & 6                                             & 204                                           \\
			Phalombe                                    & 17                                                       & 165                                                & 3                                            & 27                                           & 3                                             & 316                                           \\
			Rumphi                                      & 8                                                        & 130                                                & 6                                            & 20                                           & 12                                            & 156                                           \\
			Salima                                      & 5                                                        & 168                                                & 6                                            & 23                                           & 22                                            & 416                                           \\
			Thyolo                                      & 8                                                        & 177                                                & 4                                            & 30                                           & 12                                            & 674                                           \\
			Zomba                                       & 19                                                       & 194                                                & 9                                            & 26                                           & 79                                            & 584                                           \\ \hline
			Total                                       & 278                                                      & 4427                                               & 168                                          & 659                                          & 1409                                          & 11149                   \\\hline                     
		\end{tabular}
		\caption{Summary statistics of Malawi 2015--16 DHS data, by district. These summaries are for females aged 15--29.}
		\label{tab:summarystat}
	\end{table}
	
}

%


\clearpage
\subsection*{HIV Prevalence in Malawi Modeling}

We use penalized complexity (PC) priors \citep{simpson:etal:17} in our analyses.
PC priors facilitate intuitive hyperprior assignment. In this framework, we assign prior probability to a base/simple model and a more flexible/complex model, with the base model being favored unless otherwise indicated by the data. Kullback-Leibler divergence is used to measure the distance between the base model and the more flexible model, and deviation from the base model is penalized at a constant rate. The PC prior for a parameter $\theta$ is specified using the probability statement $\Pr(\theta > U) = \alpha$, where the user chooses the values for $U$ and $\alpha$. Using this framework, we set priors for the BYM2 overall precision parameter and mixing parameter. For the overall precision parameter in the BYM2 model, we set $U = 1$, $\alpha = 0.01$, which corresponds to a prior probability of 0.99 of having residual odds ratios smaller than 2. For the mixing parameter, we set $U = .5$, $\alpha = 2/3$, which corresponds to a 67\% chance that more than 50\% of the total variation of the random effect has spatial structure. Details for the derivations of the PC prior for the BYM2 model can be found in Appendix 2 of \cite{riebler:etal:16}.



\begin{figure}[htbp]
	\centering
	\includegraphics[width=0.6\linewidth]{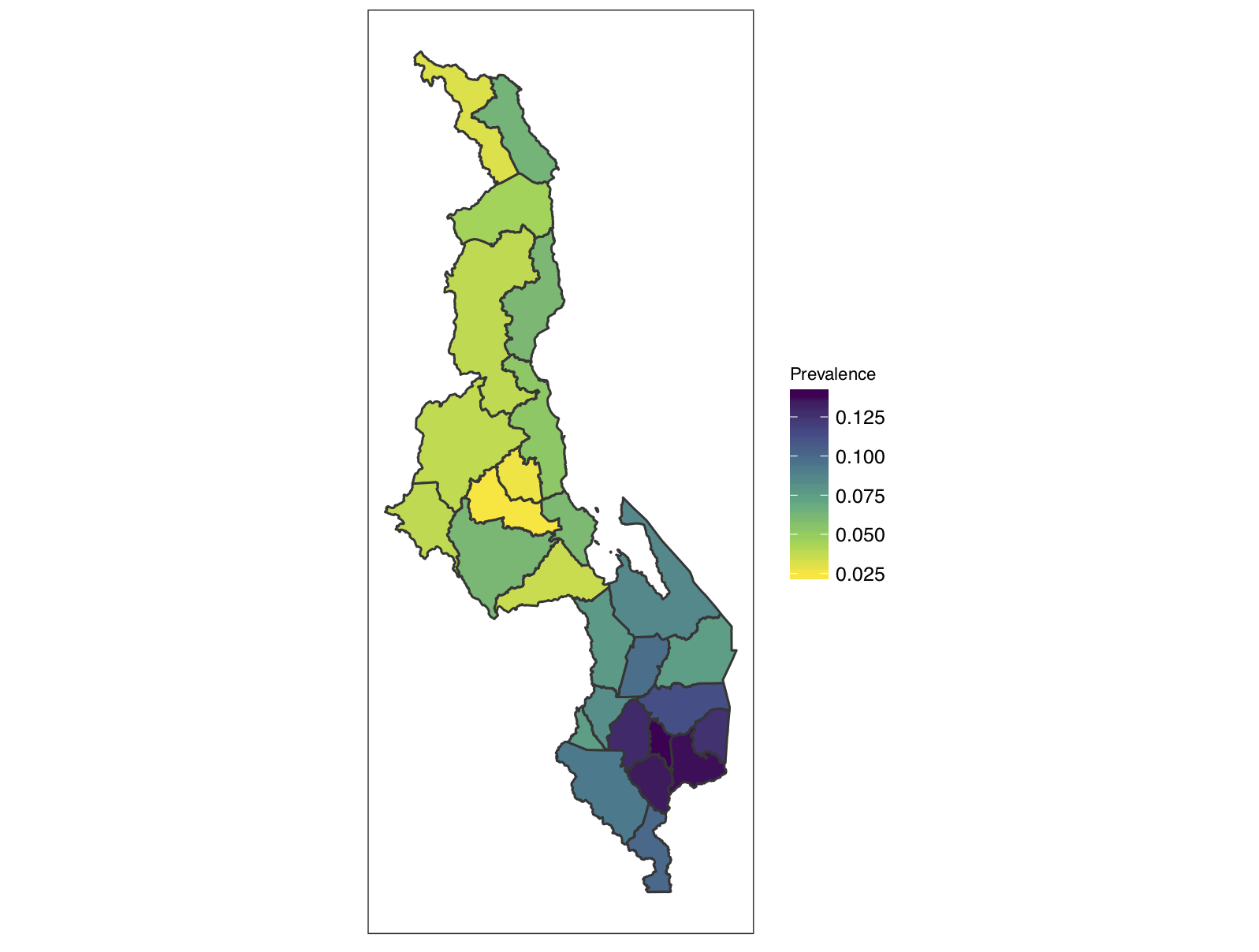}\hspace{-1in}
	\caption{Map of ANC HIV prevalence estimates.}
	\label{fig:covariates}
\end{figure}


\begin{figure}[htbp]
	\centering
	\includegraphics[width=1.0\linewidth]{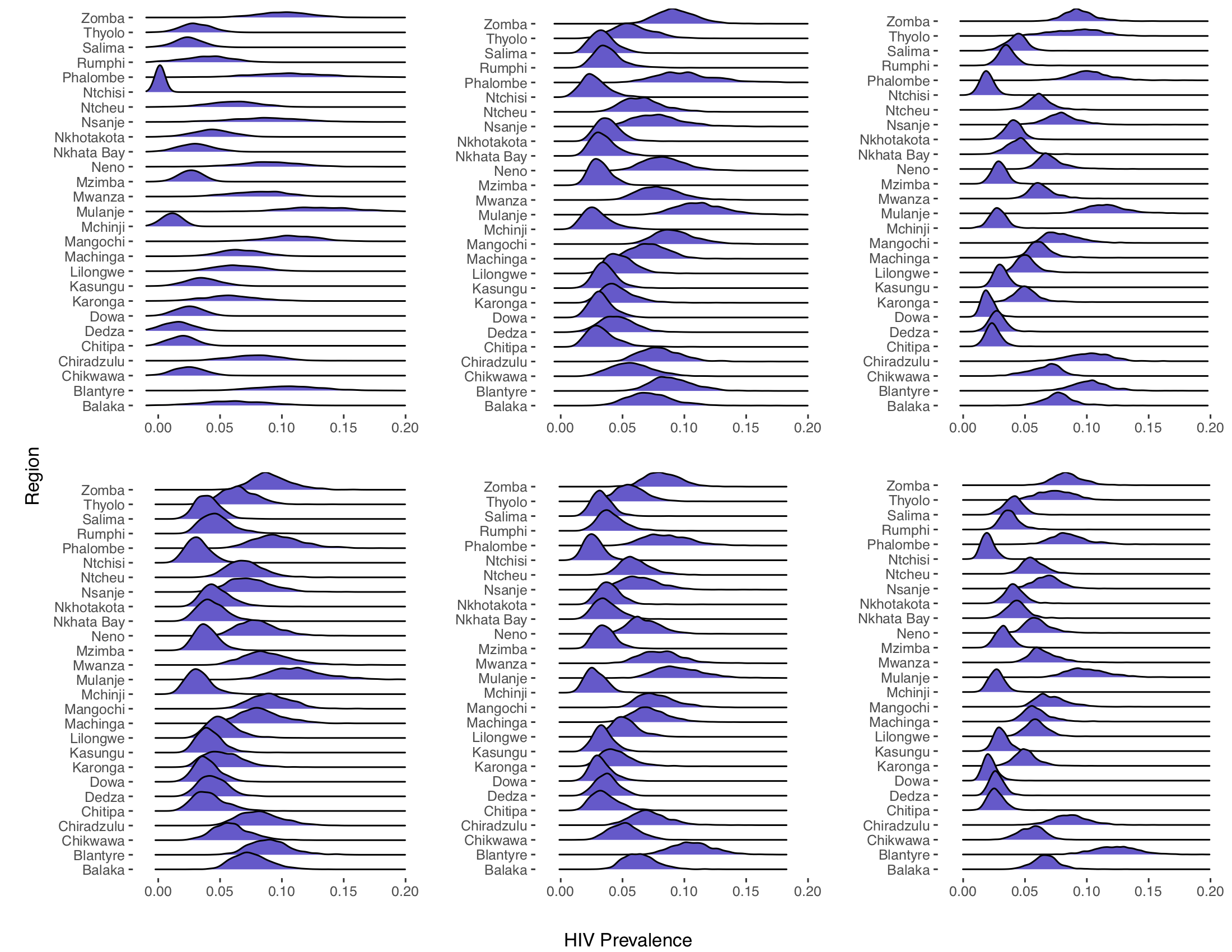}
	\caption{Posterior distributions for HIV prevalence. Top row area-level models: direct;  smoothed direct; smoothed direct with ANC covariate. Bottom row unit-level (betabinomial) models: no urban/rural, no covariate; urban/rural only; urban/rural and ANC covariate.}
	\label{fig:betabinom_ridge}
\end{figure}

\begin{figure}[htbp]
	\centering
	\includegraphics[width=0.6\linewidth]{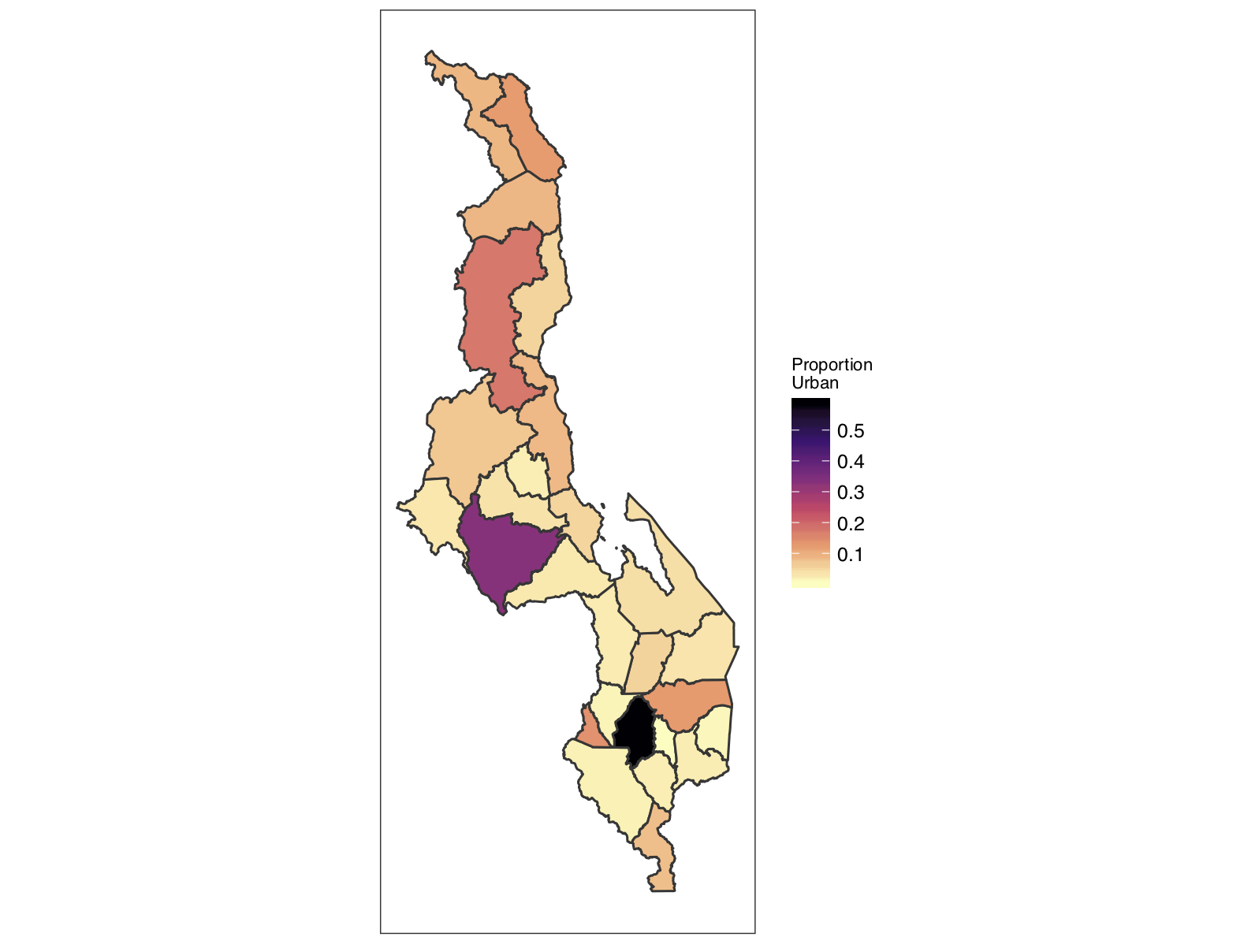}
	\caption{Map of estimate of urban proportion in each district of Malawi.}
	\label{fig:urban_plot}
\end{figure}

\begin{figure}[htbp]
	\centering
	\includegraphics[width=0.6\linewidth]{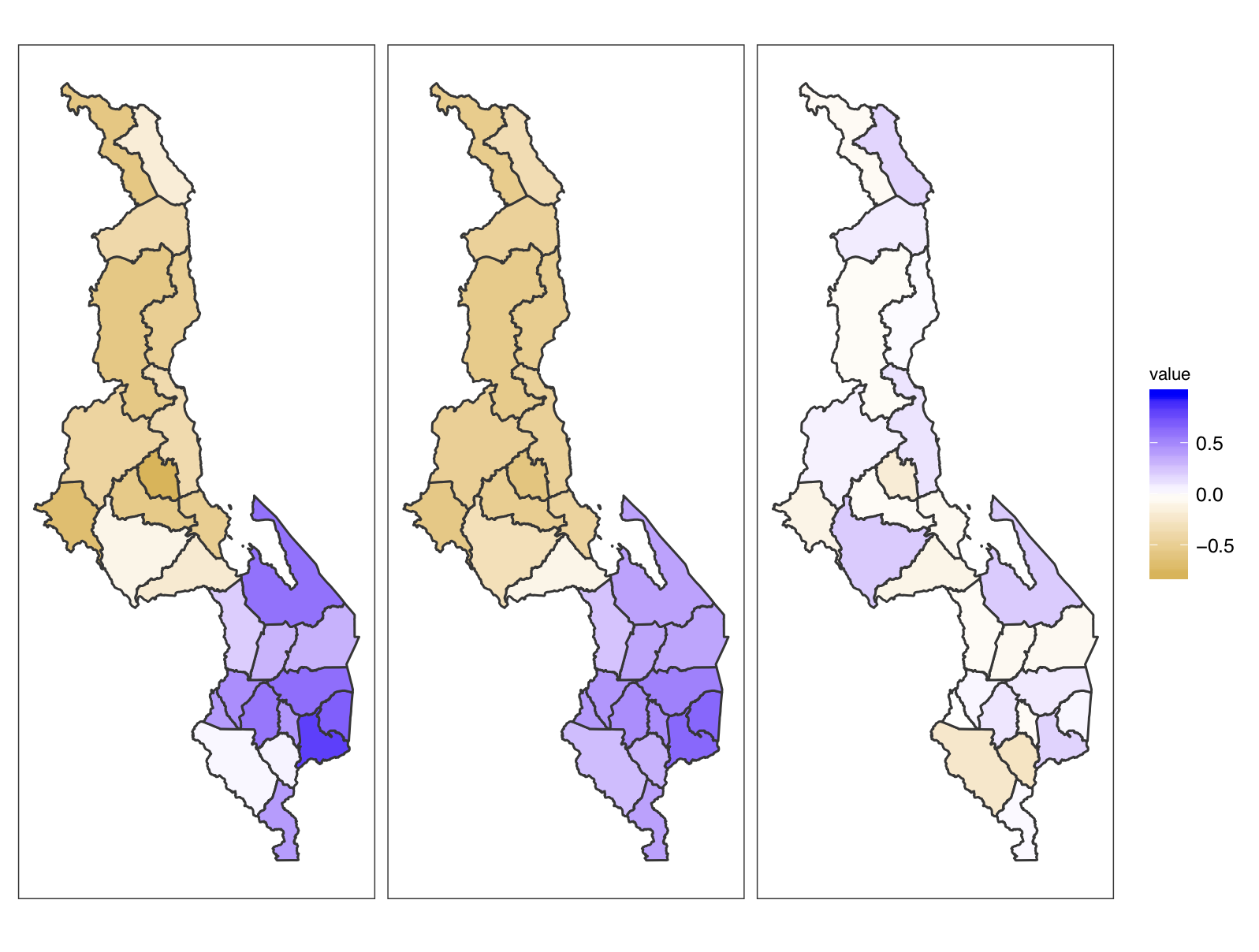}
	\caption{Random effects for the smoothed direct model with no covariate. Left: Combined. Middle: Spatial. Right: IID.}
	\label{fig:smooth_re}
\end{figure}
\begin{figure}[htbp]
	\centering
	\includegraphics[width=0.6\linewidth]{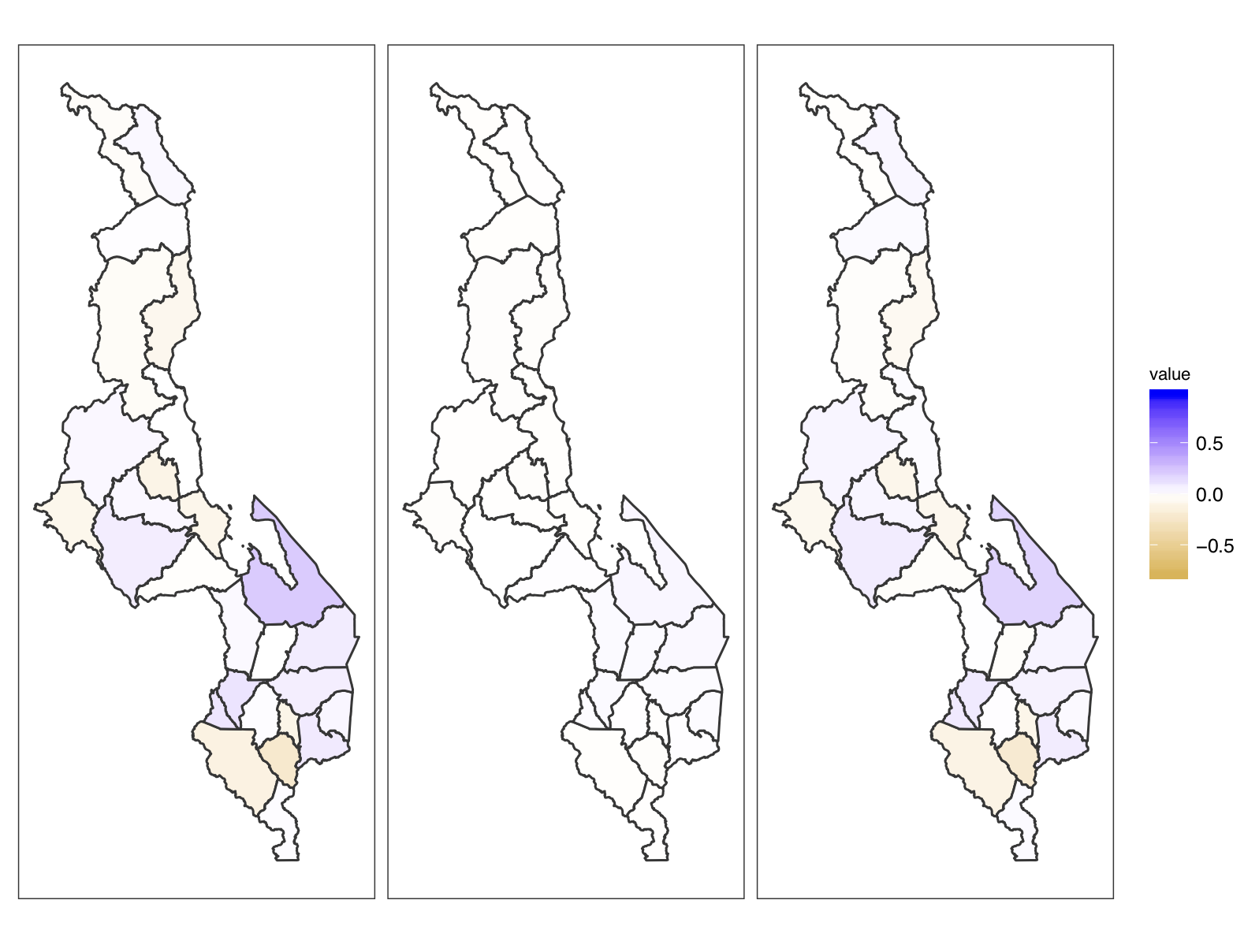}
	\caption{Random effects for the betabinomial model with ANC covariate. Left: Combined. Middle: Spatial. Right: IID}
	\label{fig:smooth_logitanc_re}
\end{figure}

\begin{figure}[htbp]
	\centering
	\includegraphics[width=0.6\linewidth]{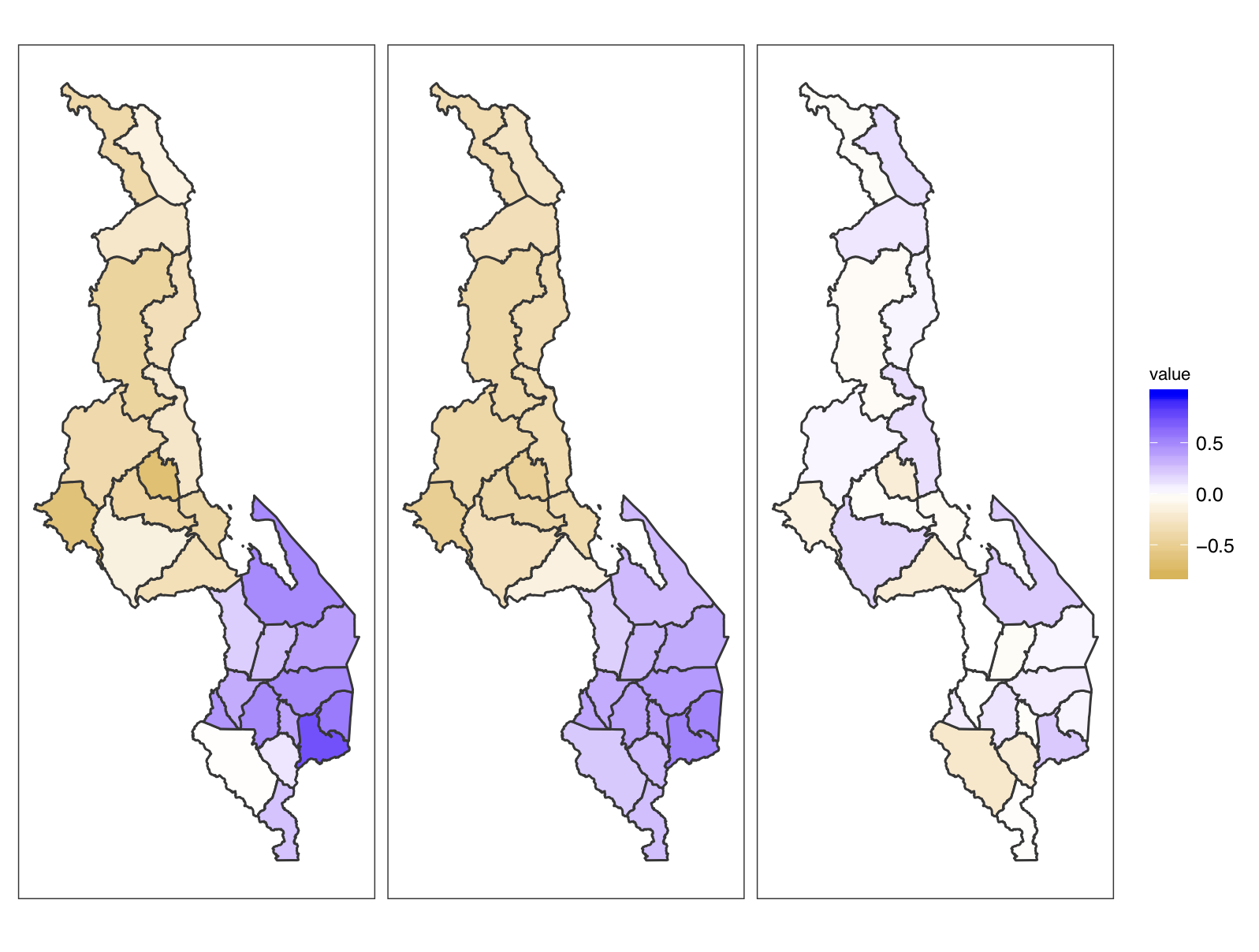}
	\caption{Random effects for the betabinomial model with no urban/rural and no covariate. Left: Combined. Middle: Spatial. Right: IID}
	\label{fig:bb_re}
\end{figure}

\begin{figure}[htbp]
	\centering
	\includegraphics[width=0.6\linewidth]{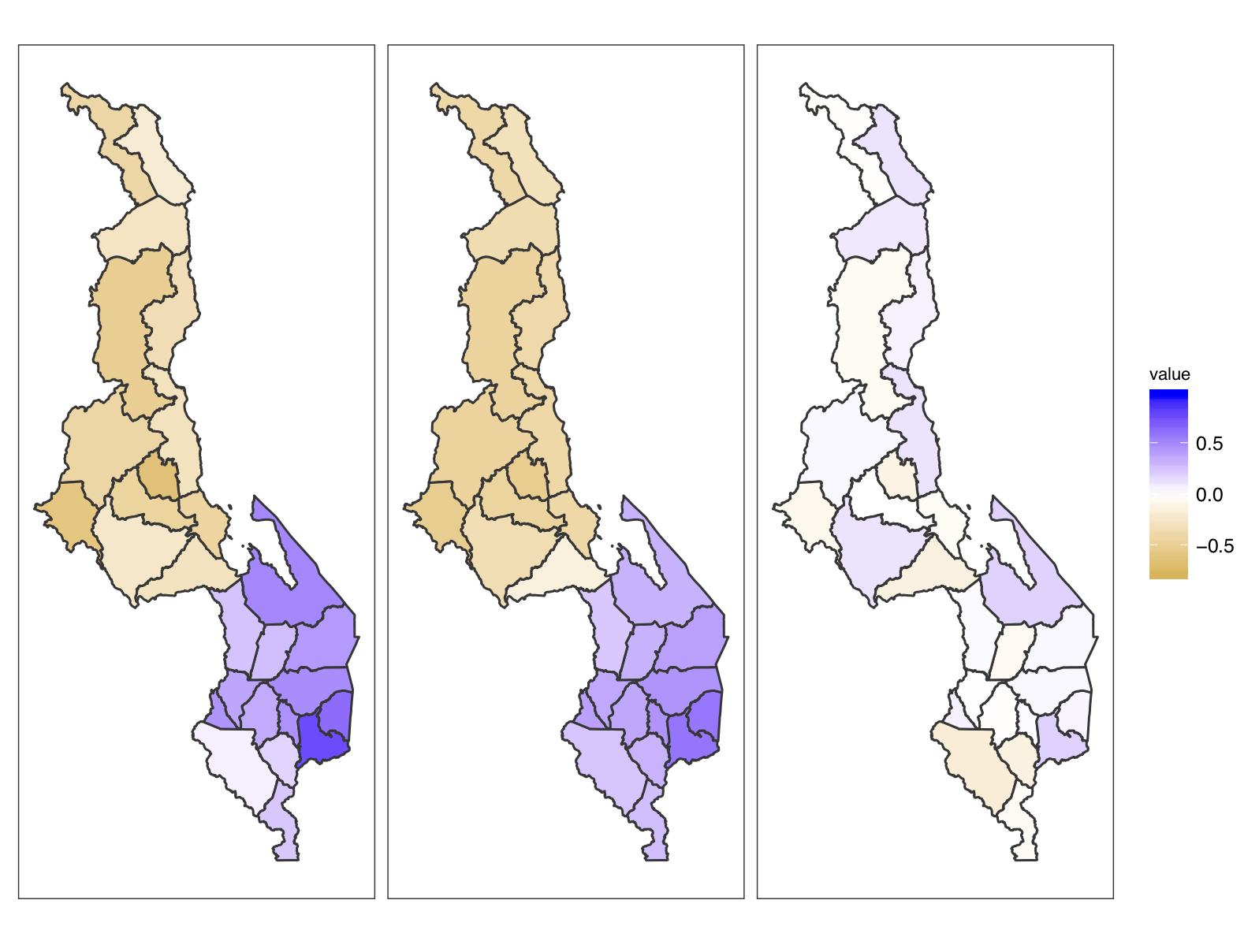}
	\caption{Random effects for the betabinomial model with urban/rural and no covariate. Left: Combined. Middle: Spatial. Right: IID}
	\label{fig:bb_strat_re}
\end{figure}

\begin{figure}[htbp]
	\centering
	\includegraphics[width=0.6\linewidth]{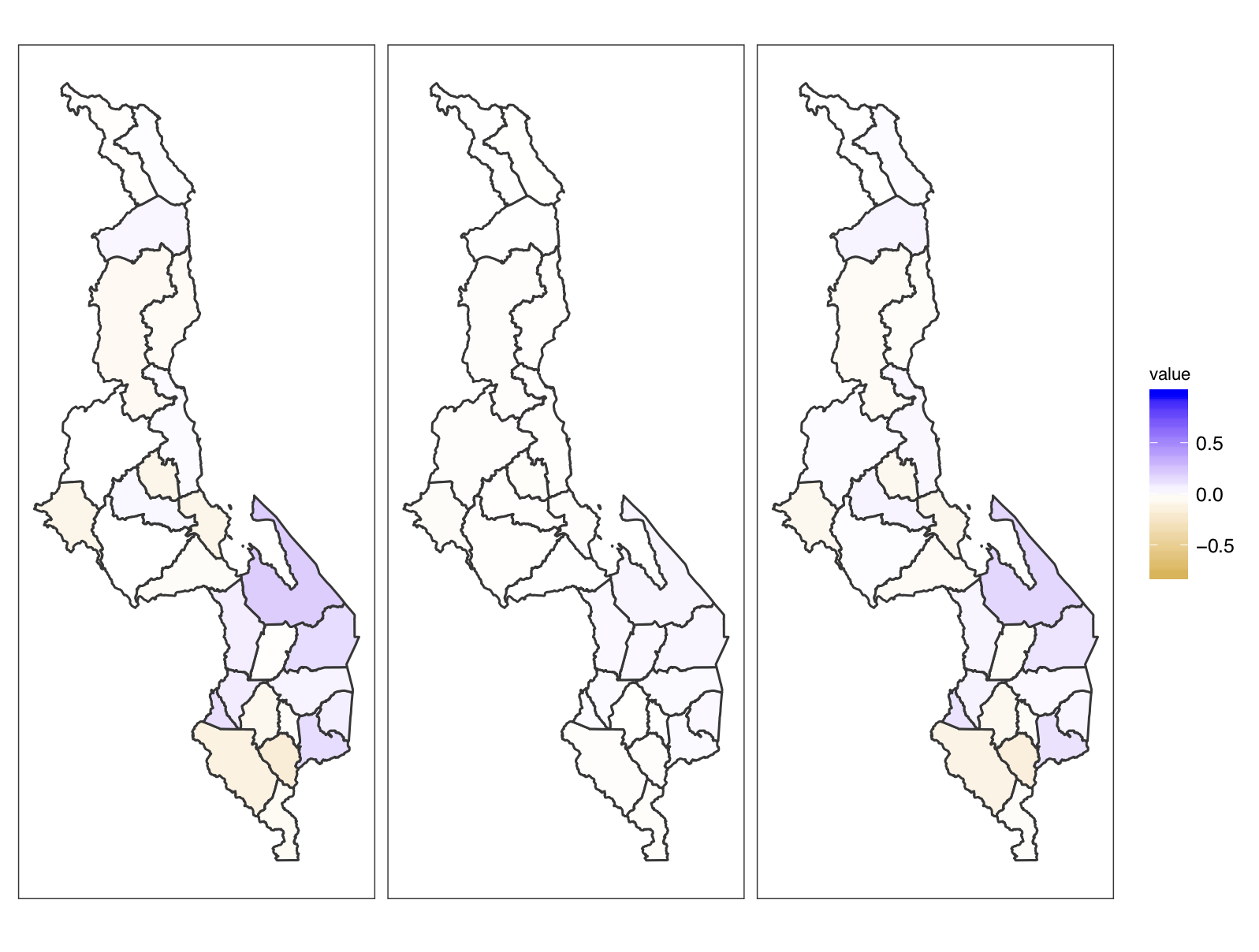}
	\caption{Random effects for the betabinomial model with urban/rural and ANC covariate. Left: Combined. Middle: Spatial. Right: IID}
	\label{fig:bb_strat_logitanc_re}
\end{figure}

\begin{figure}[tbp]
	\centering
	\includegraphics[width=1.0\linewidth]{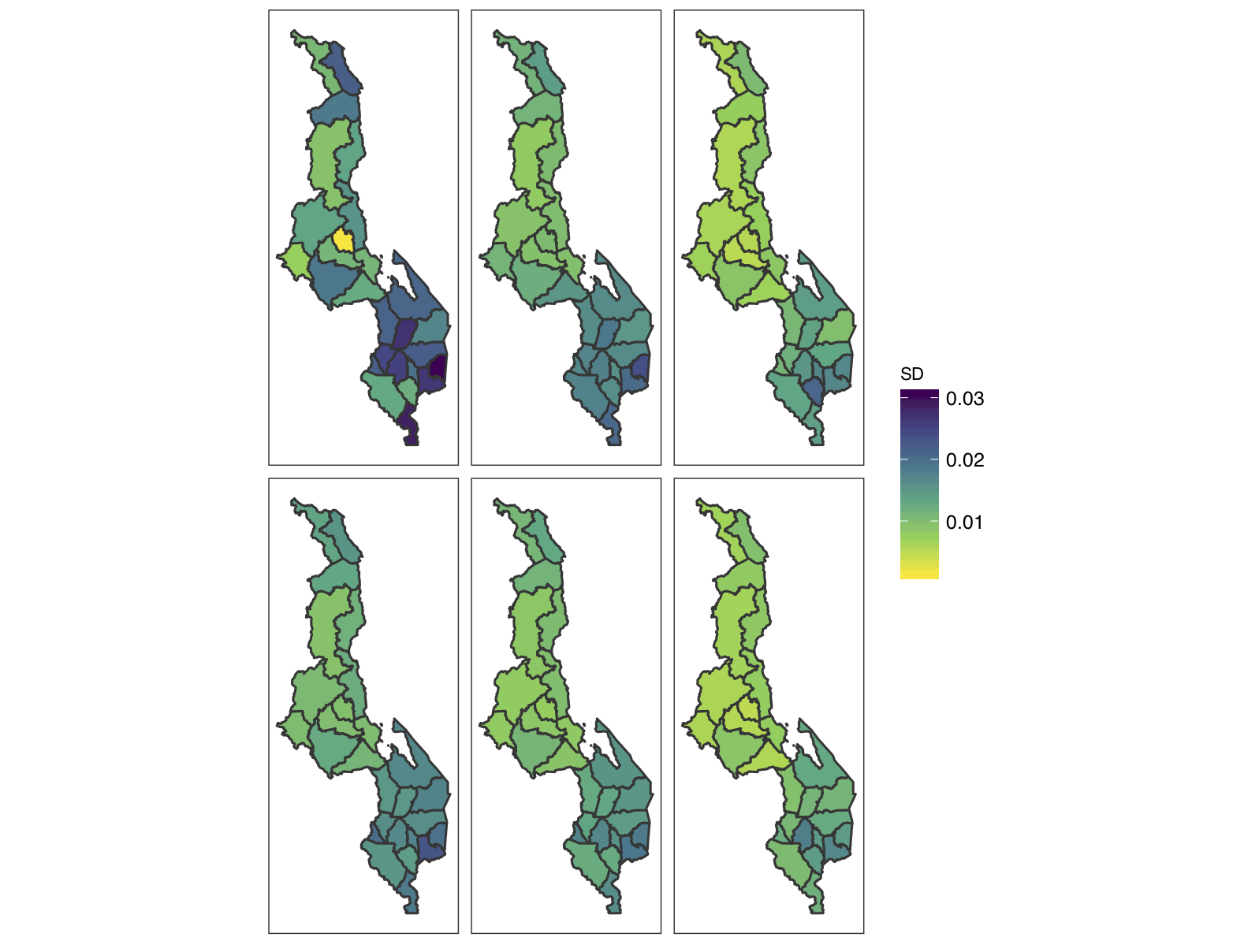}\\
	\caption{Uncertainty estimates (standard errors for direct estimates, posterior standard deviations for the remainder) of HIV prevalence among females aged 15--29 in districts of Malawi. Top row shows area-level models: direct estimates; smoothed direct estimates with no ANC covariate; smooth direct estimates with ANC covariate. Bottom row shows unit-level models: no urban/rural adjustment, no ANC covariate; urban/rural adjusted, no ANC covariate; urban/rural covariate and ANC covariate.}
	\label{fig:Malawi2}
\end{figure}

\begin{figure}[tbp]
	\centering
	\includegraphics[width=1.0\linewidth]{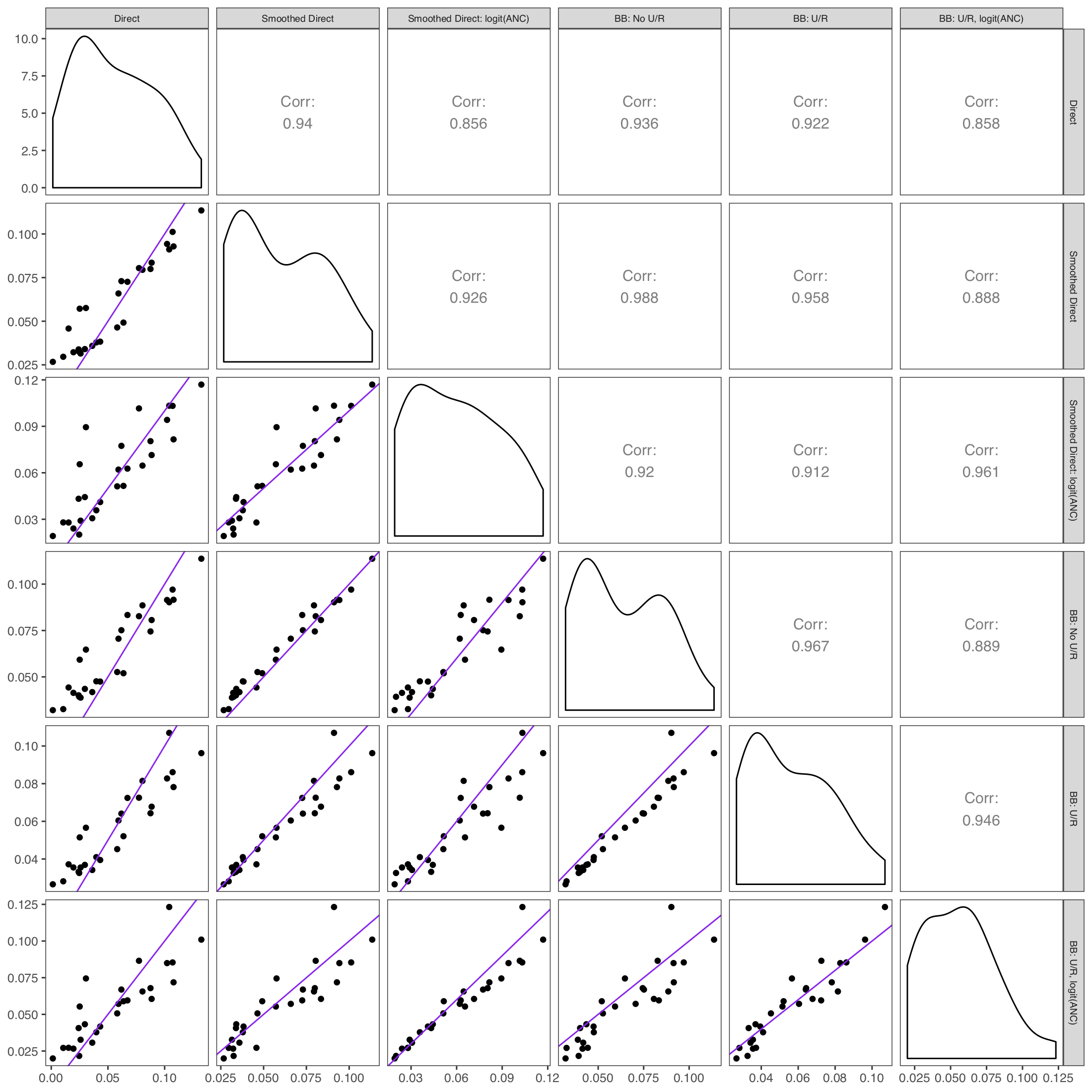}\\
	\caption{Estimates (weighted estimates for direct estimation and posterior medians for the remainder) of HIV prevalence in 27 districts of Malawi, under six estimation methods. The purple lines have intercept 0 and slope 1.}
	\label{fig:model_comparison6_mean}
\end{figure}

\begin{figure}[tbp]
	\centering
	\includegraphics[width=1.0\linewidth]{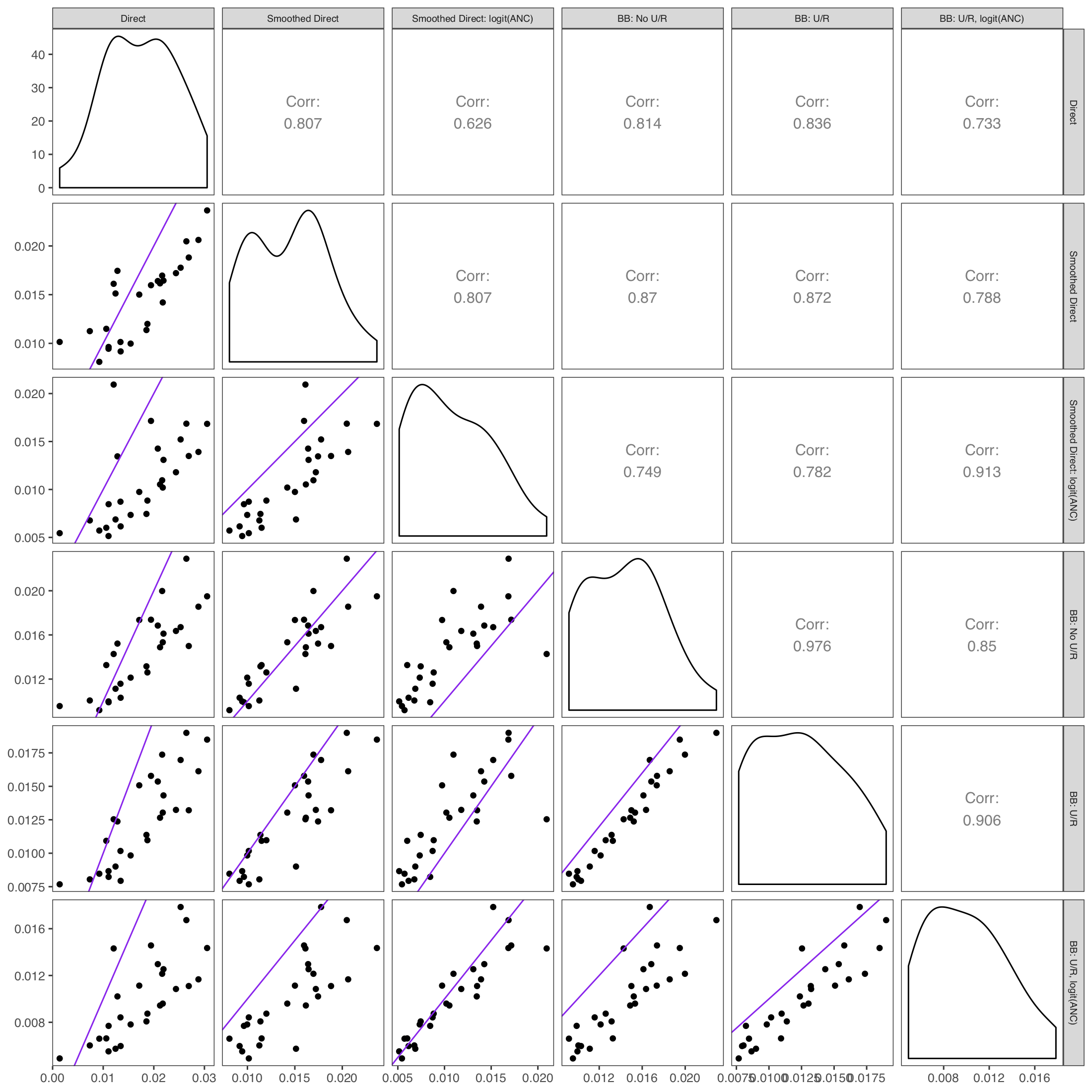}\\
	\caption{Uncertainty estimates (standard errors for direct estimation, posterior standard deviations for the remainder) of HIV prevalence in 27 districts of Malawi, under six estimation methods. The purple lines have intercept 0 and slope 1.}
	\label{fig:model_comparison6_sd}
\end{figure}

\begin{figure}[tbp]
	\centering
	\includegraphics[width=0.6\linewidth]{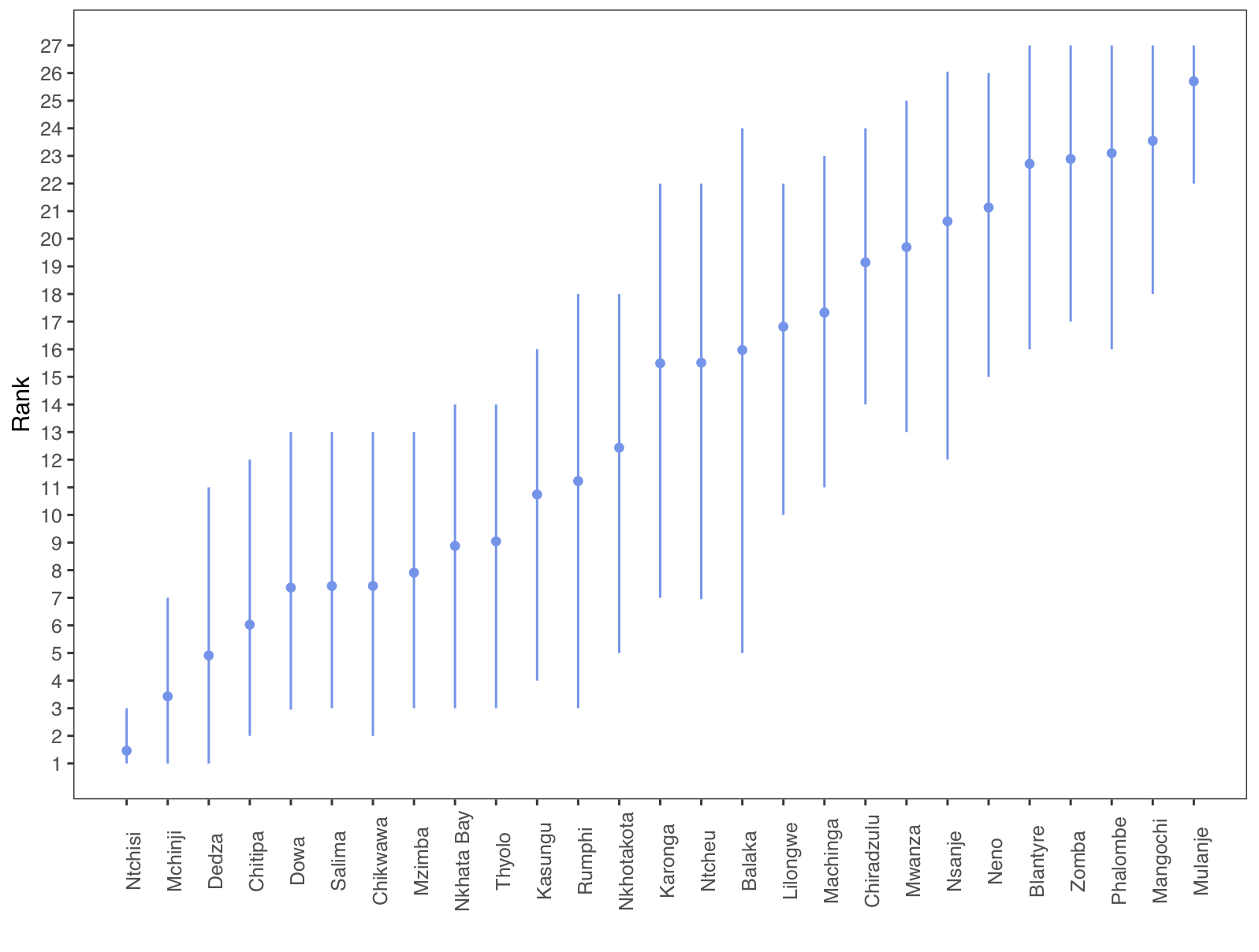}\\
	\caption{Distributions on the rankings for the direct estimates. The lines represent 90\% intervals based on samples from the posterior, with rank = 1 on the y-axis corresponding to the lowest HIV prevalence and rank = 27 corresponding to the highest HIV prevalence.}
	\label{fig:direct_rank_plot}
\end{figure}

\begin{figure}[tbp]
	\centering
	\includegraphics[width=0.6\linewidth]{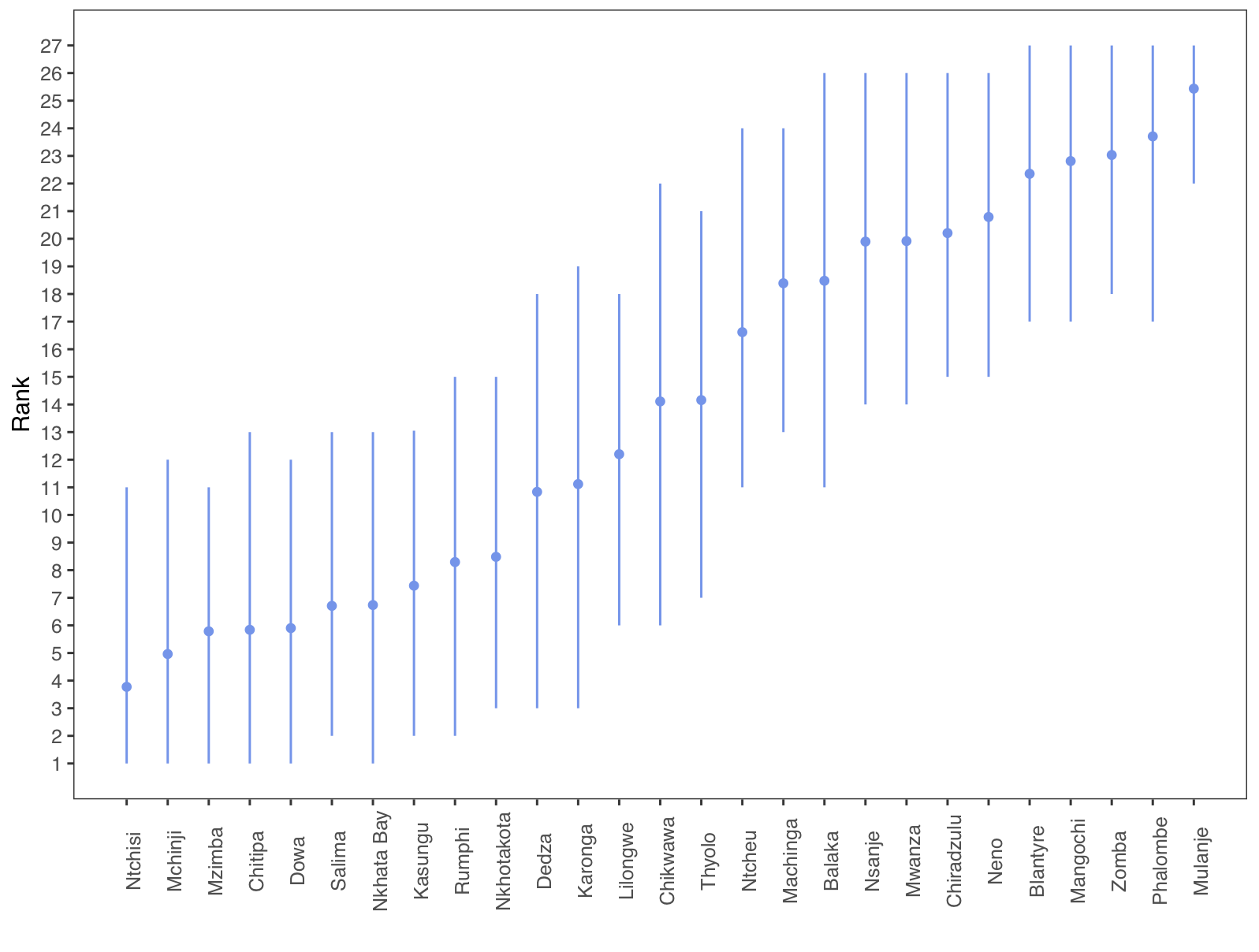}\\
	\caption{Distributions on the rankings for the smoothed direct estimates. The lines represent 90\% intervals based on samples from the posterior, with rank = 1 on the y-axis corresponding to the lowest HIV prevalence and rank = 27 corresponding to the highest HIV prevalence.}
	\label{fig:smooth_rank_plot}
\end{figure}

\begin{figure}[tbp]
	\centering
	\includegraphics[width=0.6\linewidth]{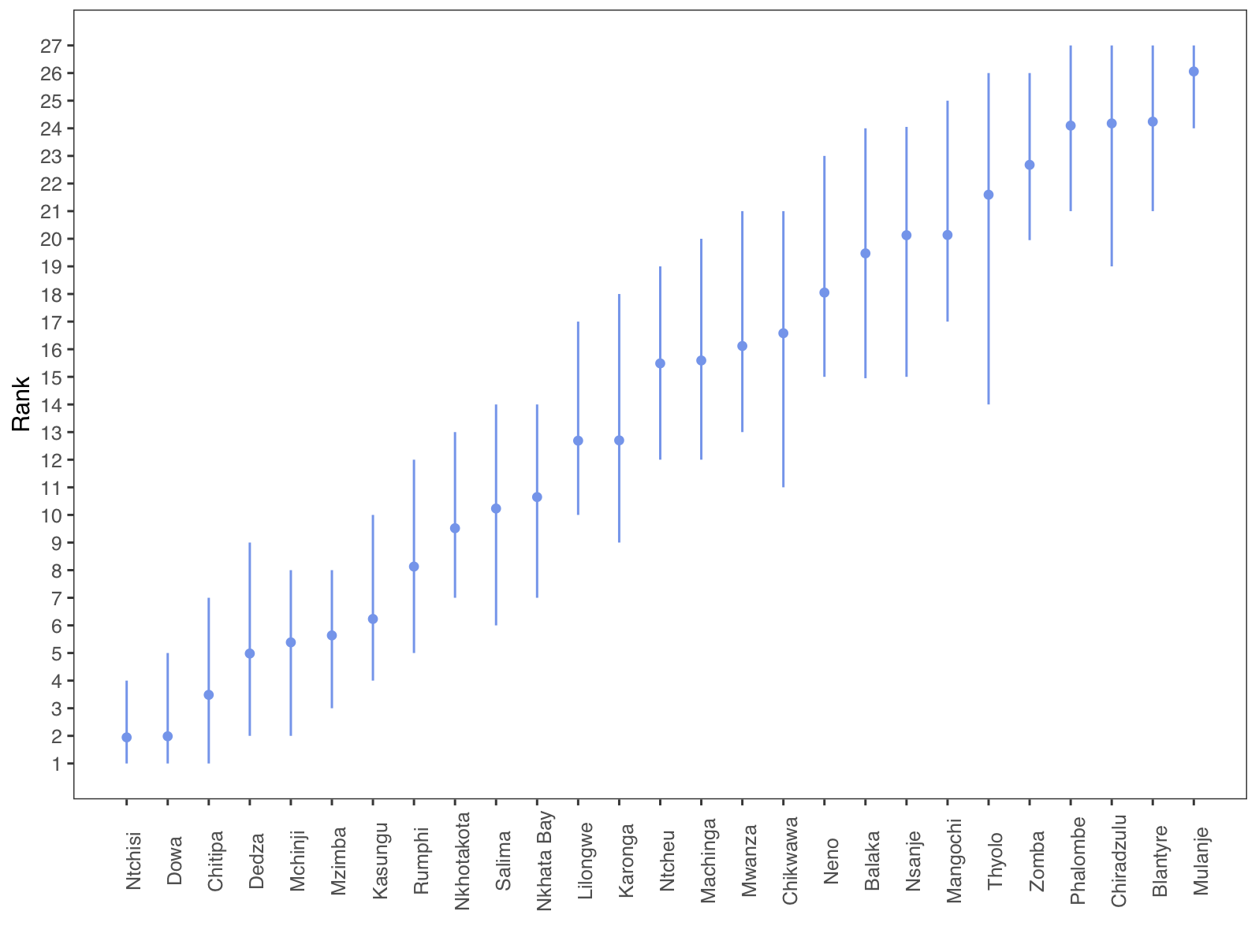}\\
	\caption{Distributions on the rankings for the smoothed direct estimates with the ANC covariate. The lines represent 90\% intervals based on samples from the posterior, with rank = 1 on the y-axis corresponding to the lowest HIV prevalence and rank = 27 corresponding to the highest HIV prevalence.}
	\label{fig:smooth_logitanc_rank_plot}
\end{figure}

\begin{figure}[tbp]
	\centering
	\includegraphics[width=0.6\linewidth]{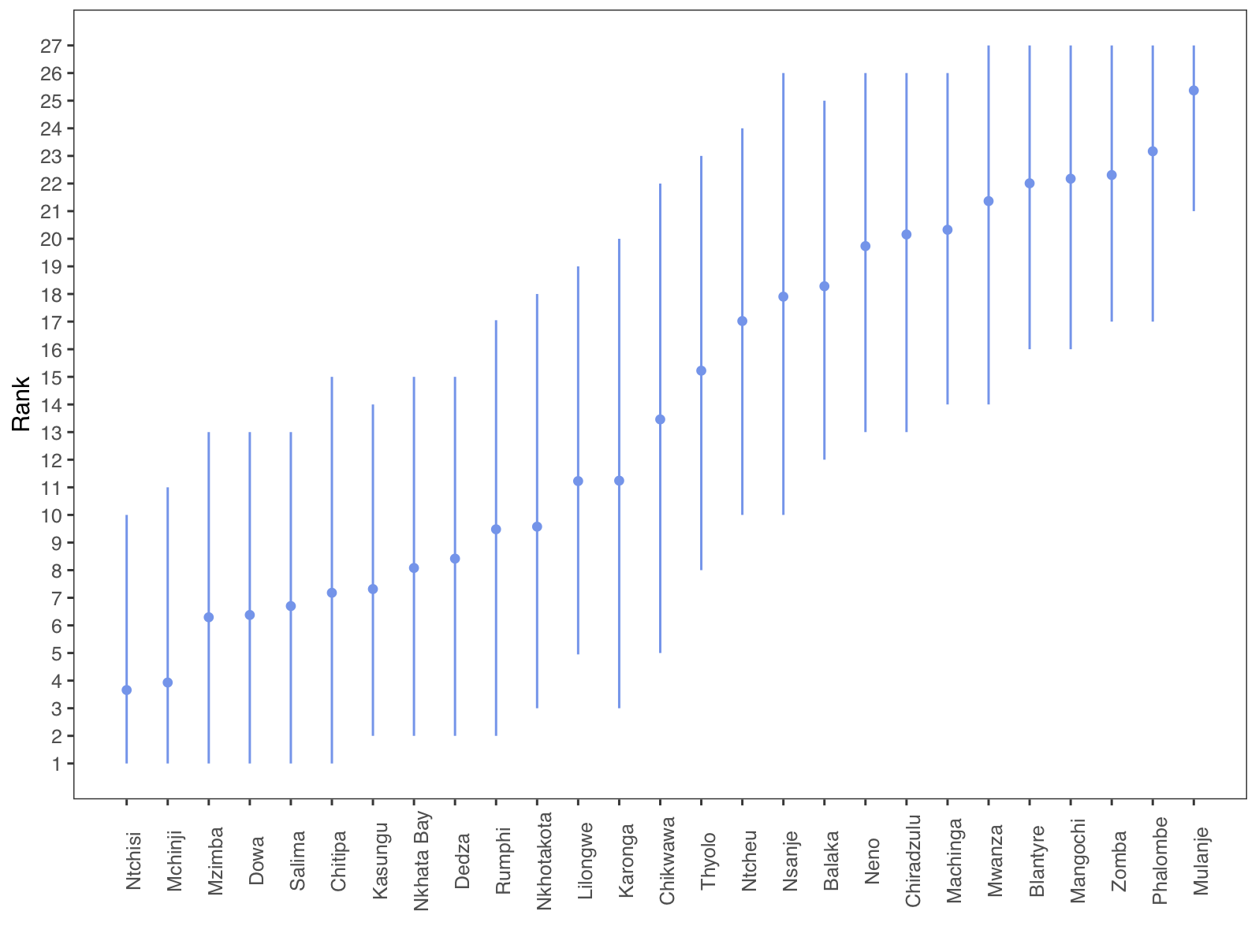}\\
	\caption{Distributions on the rankings for the betabinomial unit-level model with no adjustment for urban/rural and no covariate. The lines represent 90\% intervals based on samples from the posterior, with rank = 1 on the y-axis corresponding to the lowest HIV prevalence and rank = 27 corresponding to the highest HIV prevalence.}
	\label{fig:bbmod_rank_plot}
\end{figure}

\begin{figure}[tbp]
	\centering
	\includegraphics[width=0.6\linewidth]{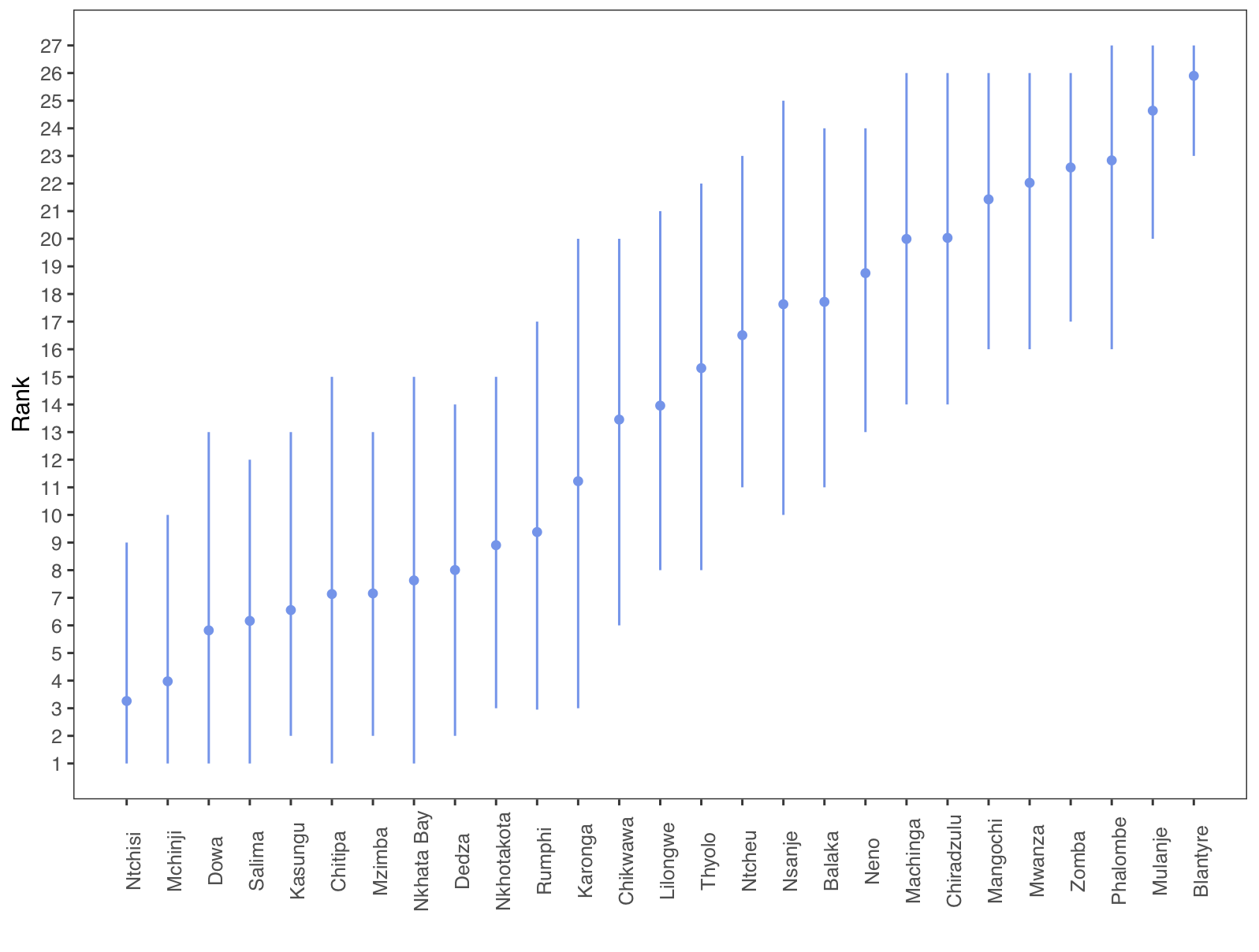}\\
	\caption{Distributions on the rankings for the betabinomial unit-level model with adjustment for urban/rural and no covariate. The lines represent 90\% intervals based on samples from the posterior, with rank = 1 on the y-axis corresponding to the lowest HIV prevalence and rank = 27 corresponding to the highest HIV prevalence.}
	\label{fig:bbmod_strat_rank_plot}
\end{figure}

\begin{figure}[tbp]
	\centering
	\includegraphics[width=0.6\linewidth]{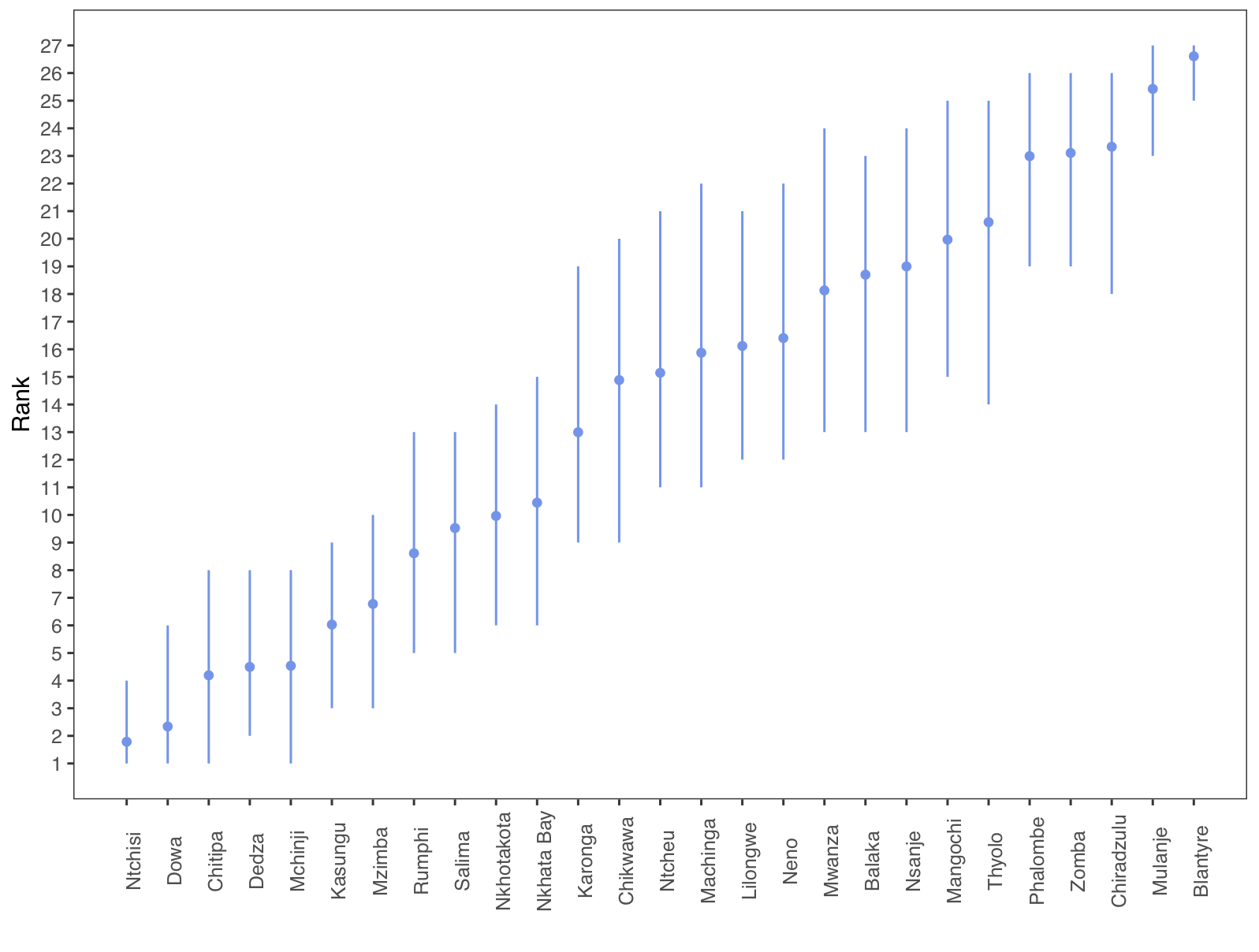}\\
	\caption{Distributions on the rankings for the betabinomial unit-level model with adjustment for urban/rural and ANC covariate. The lines represent 90\% intervals based on samples from the posterior, with rank = 1 on the y-axis corresponding to the lowest HIV prevalence and rank = 27 corresponding to the highest HIV prevalence.}
	\label{fig:bbmod_strat_logitanc_rank_plot}
\end{figure}

\clearpage
\subsection*{Model Assessment}

\begin{figure}[htbp]
	\centering
	\includegraphics[width=0.3\linewidth]{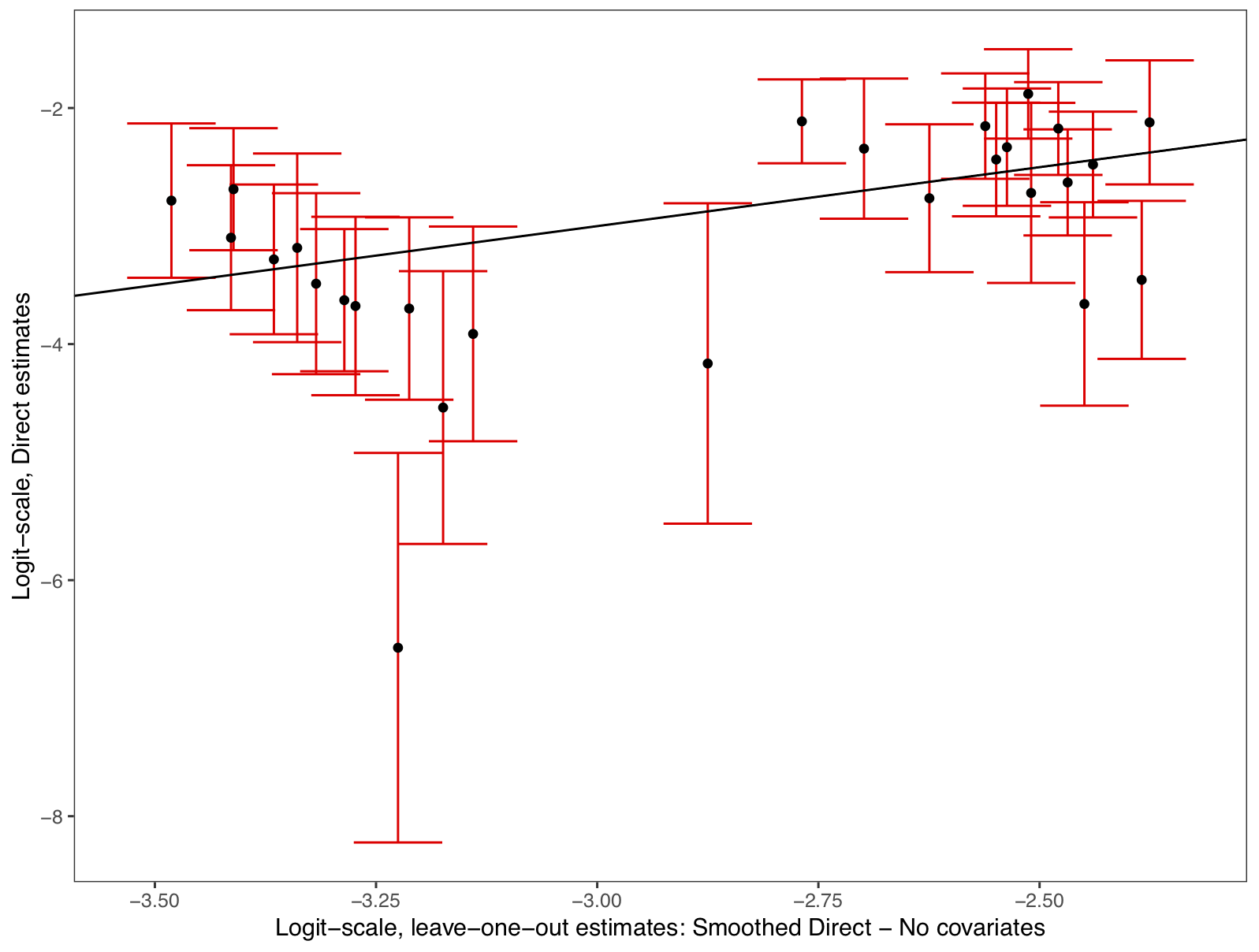}
	\includegraphics[width=0.3\linewidth]{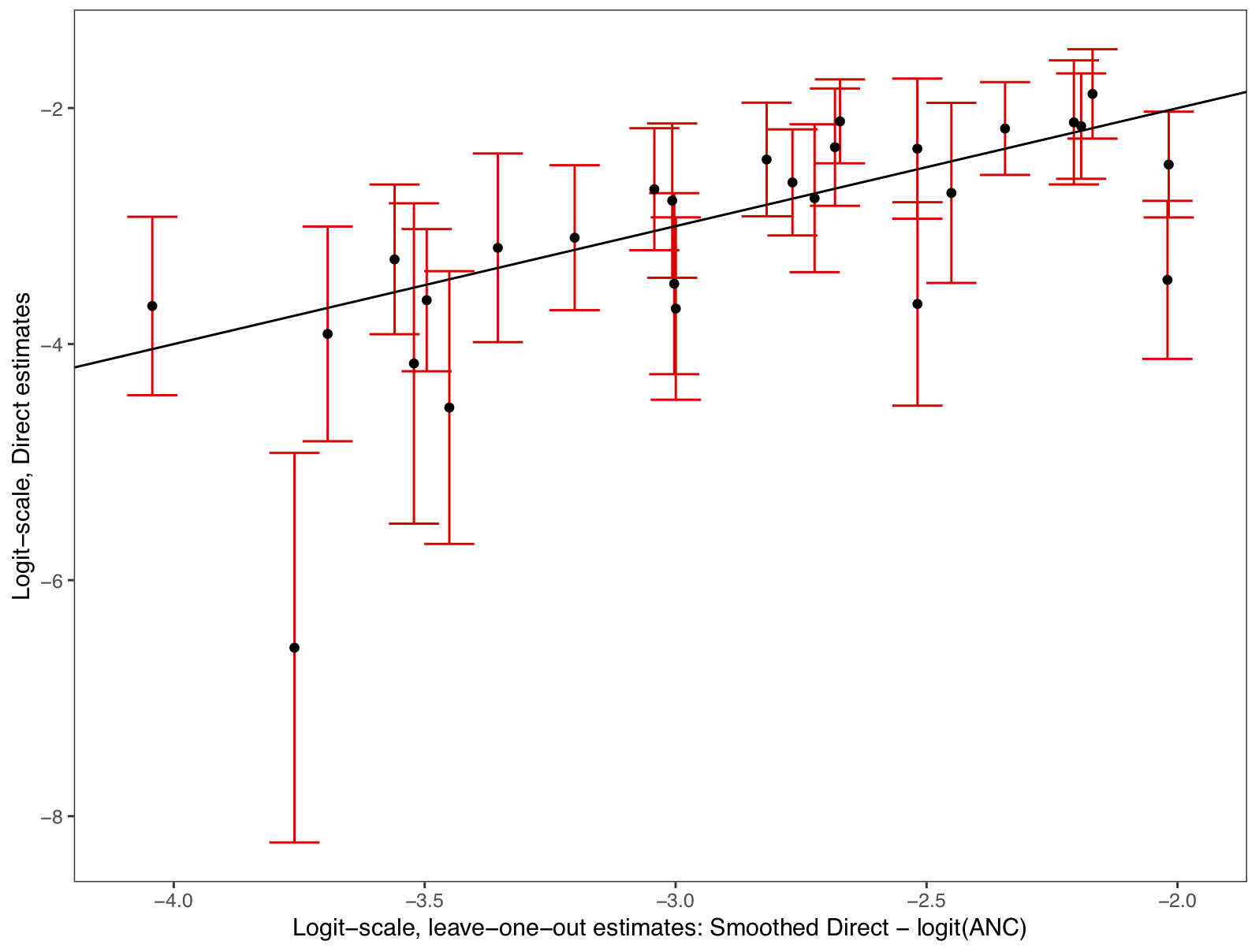}\\
	\includegraphics[width=0.3\linewidth]{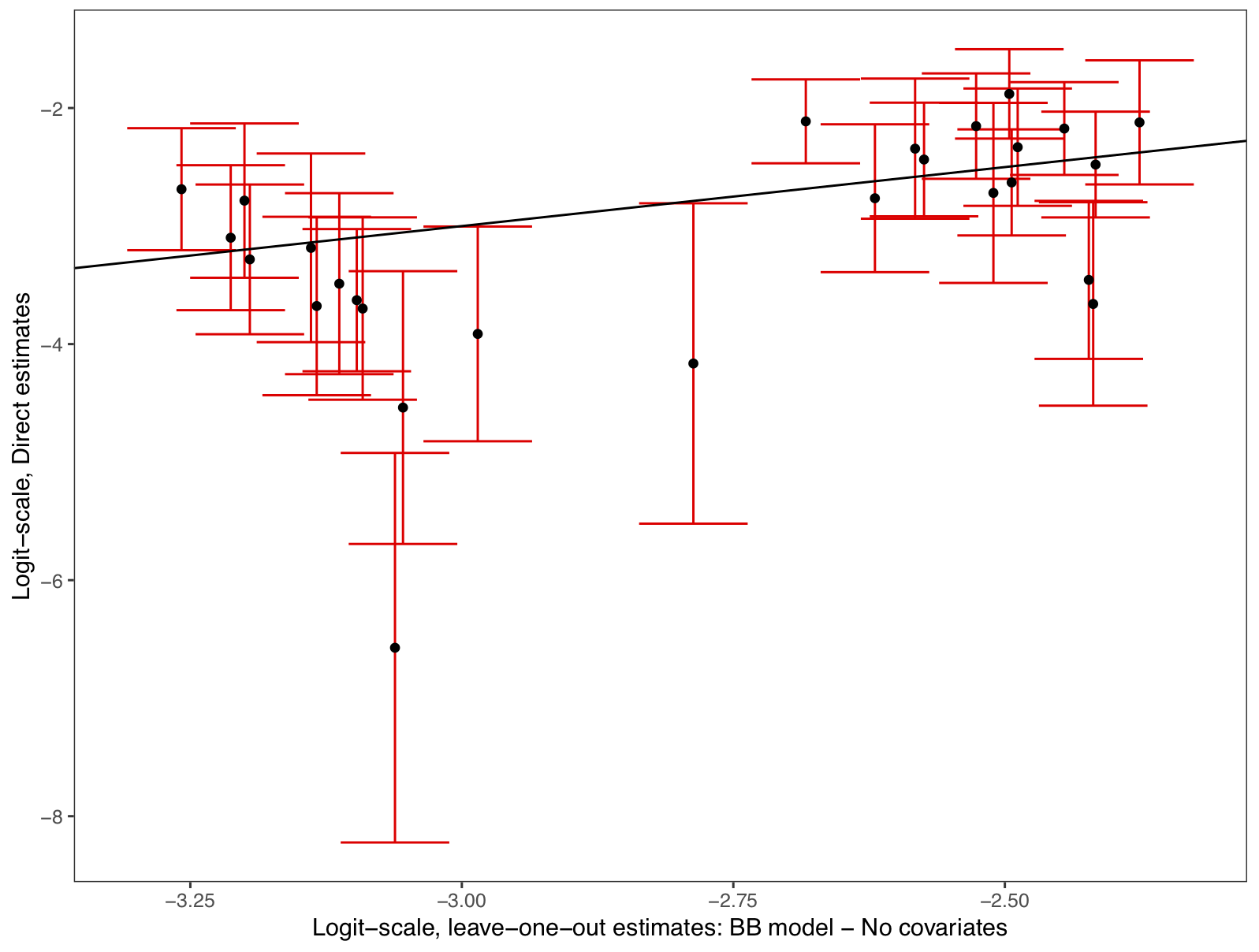}
	\includegraphics[width=0.3\linewidth]{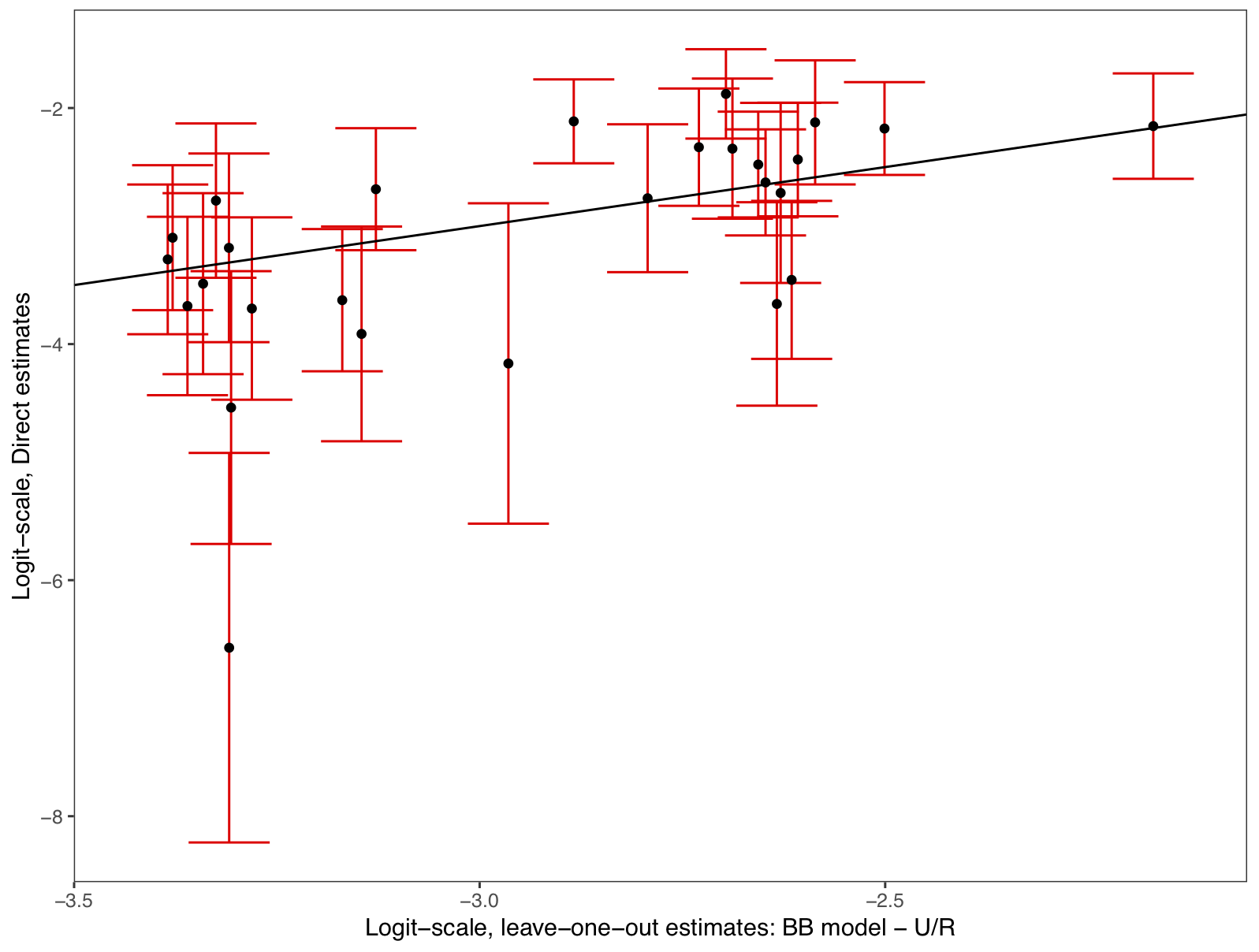}\\
	\includegraphics[width=0.3\linewidth]{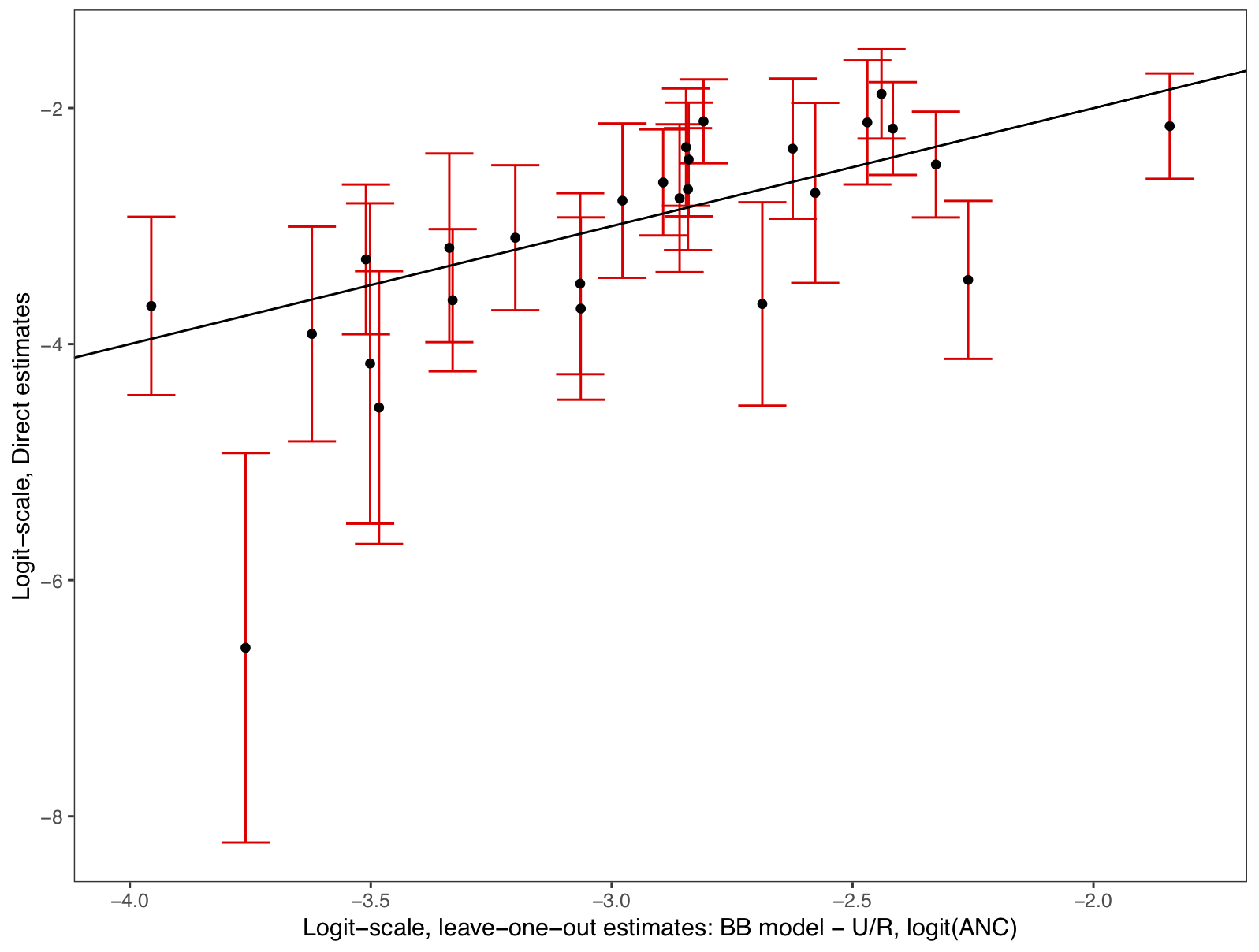}
	\caption{Leave-one cross-validation predictions. In each plot we have the direct estimates (with uncertainty bars) plotted against the posterior medians of the predictive distribution. Top row: Smoothed direct, no ANC covariate; smoothed direct, ANC covariate. Middle row: Betabinomial model with no urban/rural and no ANC covariate; betabinomial model with urban/rural but no ANC covariate. Bottom row: Betabinomial model with urban/rural and ANC covariate.}
	\label{fig:cross_validation}
\end{figure}

\clearpage
\subsection*{Admin-3 Modeling}

\begin{figure}[htbp]
	\centering
	\includegraphics[width=0.5\linewidth]{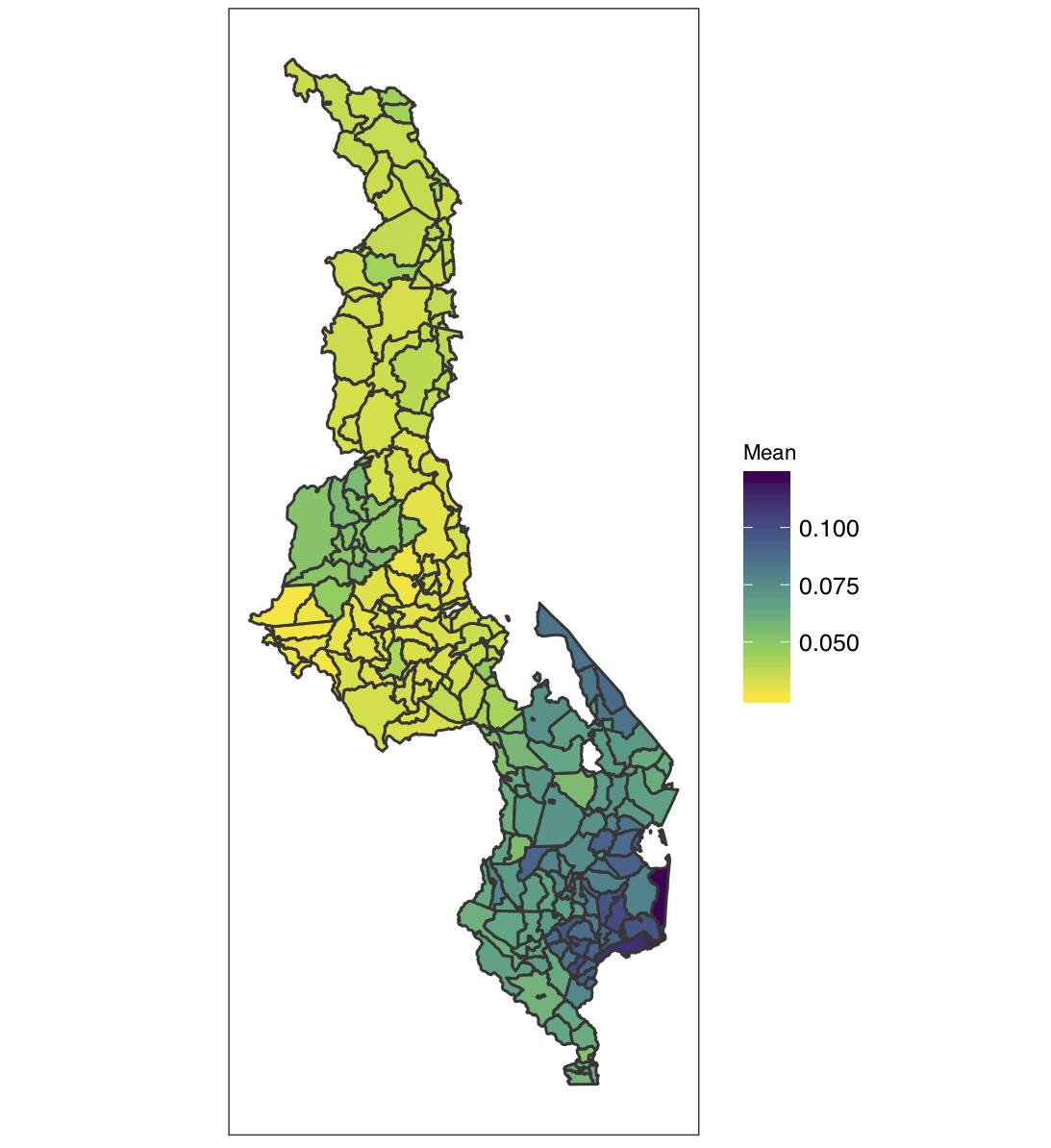}\hspace{-0.7in}
	\includegraphics[width=0.5\linewidth]{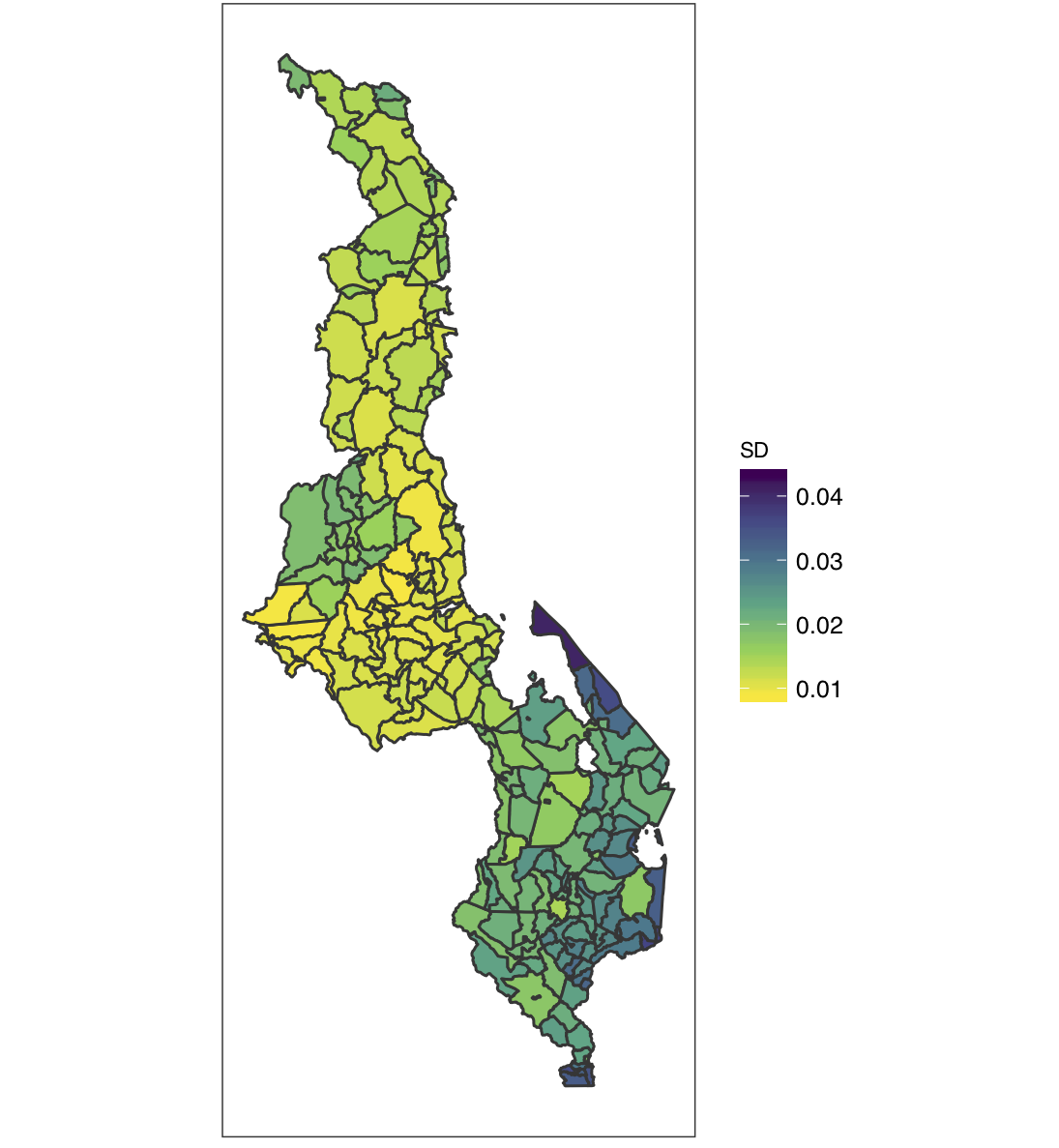}
	\caption{HIV prevalence estimates at the admin-3 level, using the betabinomial model. Left: point estimates (posterior mean). Right: posterior standard deviation.}
	\label{fig:admin3}
\end{figure}

\clearpage
\subsection*{Continuous Spatial Model Results}

The posterior median of the spatial range parameter parameter was 0.57$^\circ$, so quite large (Malawi ranges in latitude between -16.9$^\circ$ to -9.7$^\circ$ and longitude from 32.9$^\circ$ to 35.5$^\circ$), though the 95\% interval was 0.32 to 0.98, so there is a lot of uncertainty. The cluster level odds ratio was estimated as 2.3 (with a 95\% interval of 1.7,3.0).

\begin{figure}[htbp]
	\centering
	\includegraphics[width=0.5\linewidth]{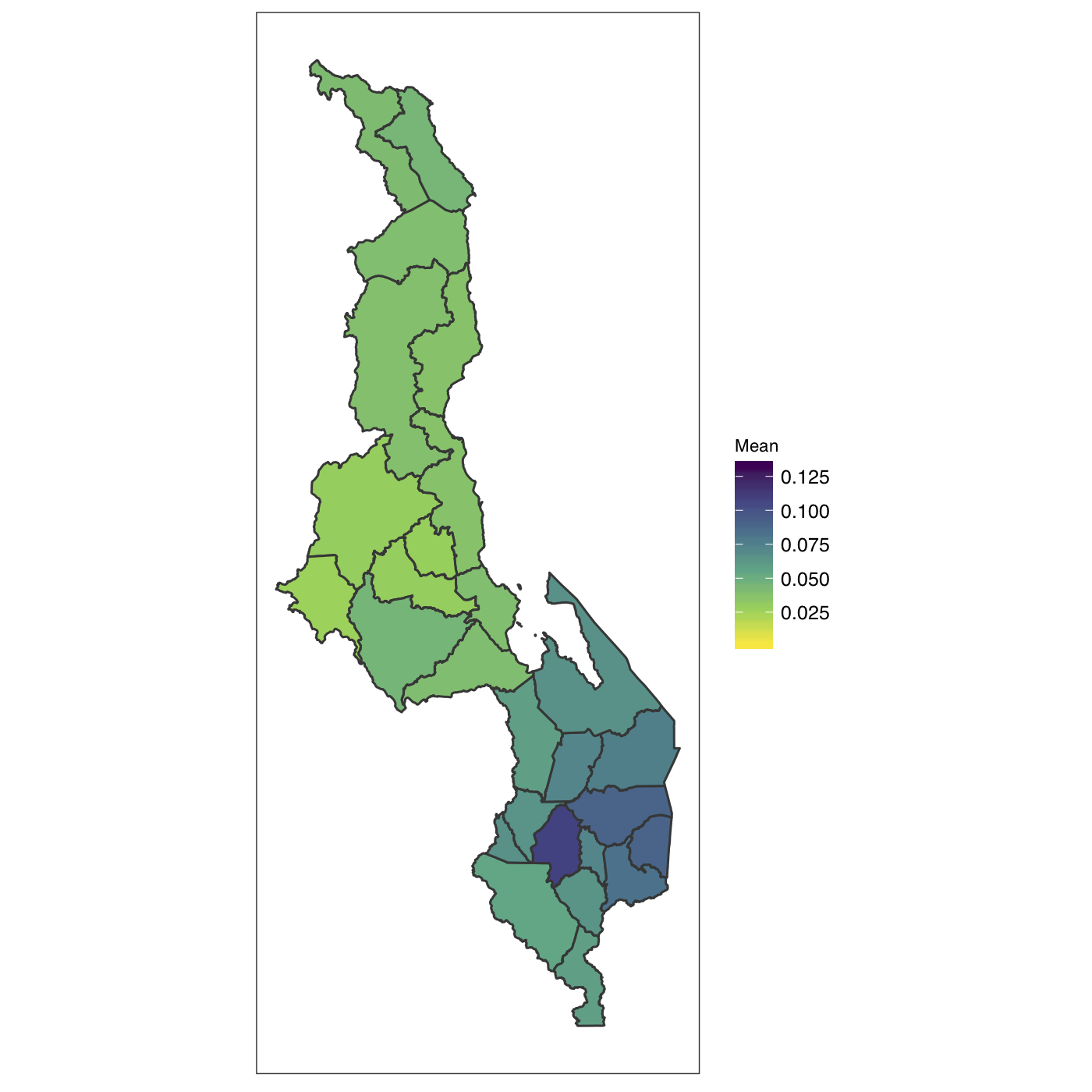}\hspace{-0.9in}
	\includegraphics[width=0.5\linewidth]{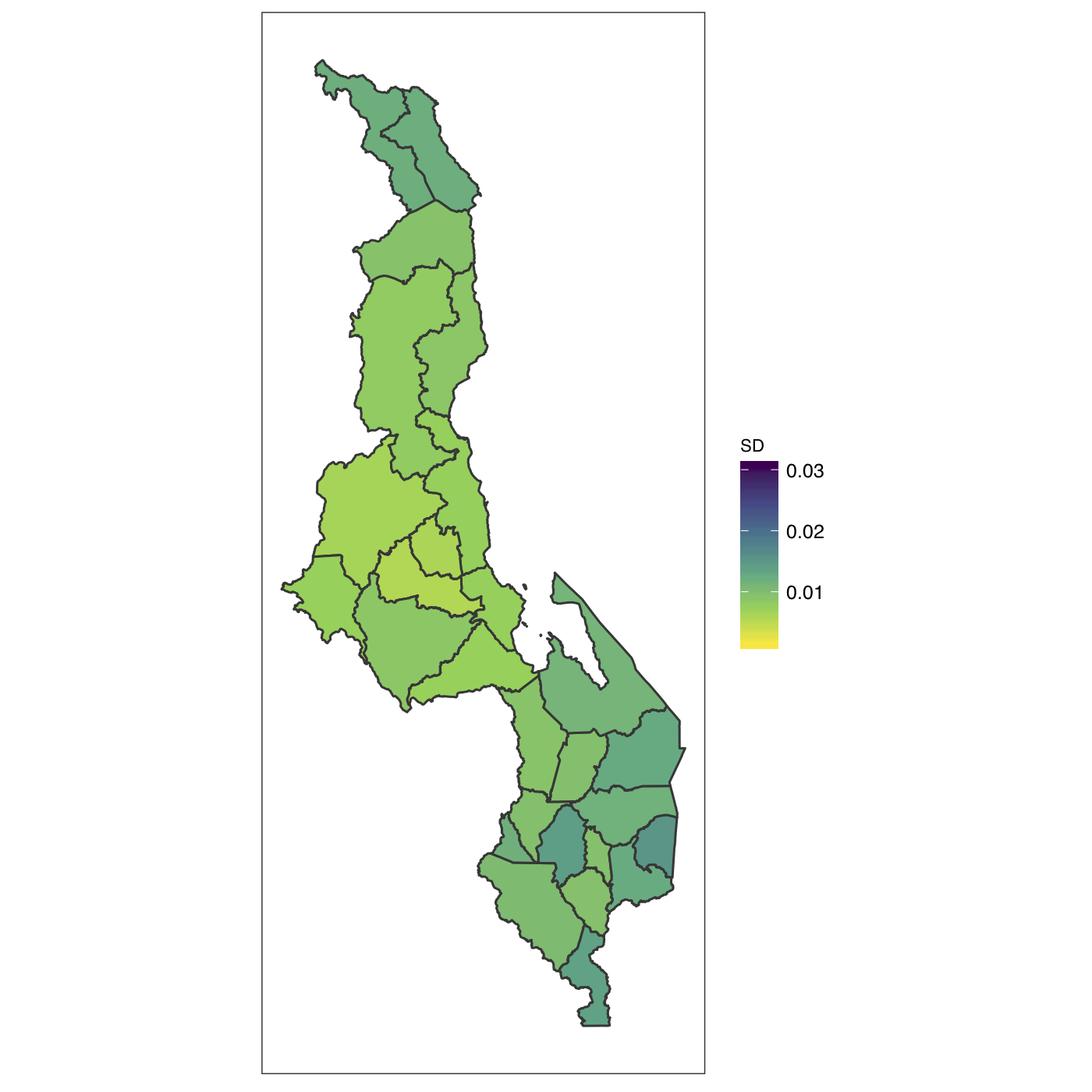}
	\caption{HIV prevalence estimates at the admin-2 level, using the SPDE model and no covariate. Left: point estimates (posterior mean). Right: posterior standard deviation.}
	\label{fig:SPDE}
\end{figure}

\begin{figure}[htbp]
	\centering
	\includegraphics[width=0.5\linewidth]{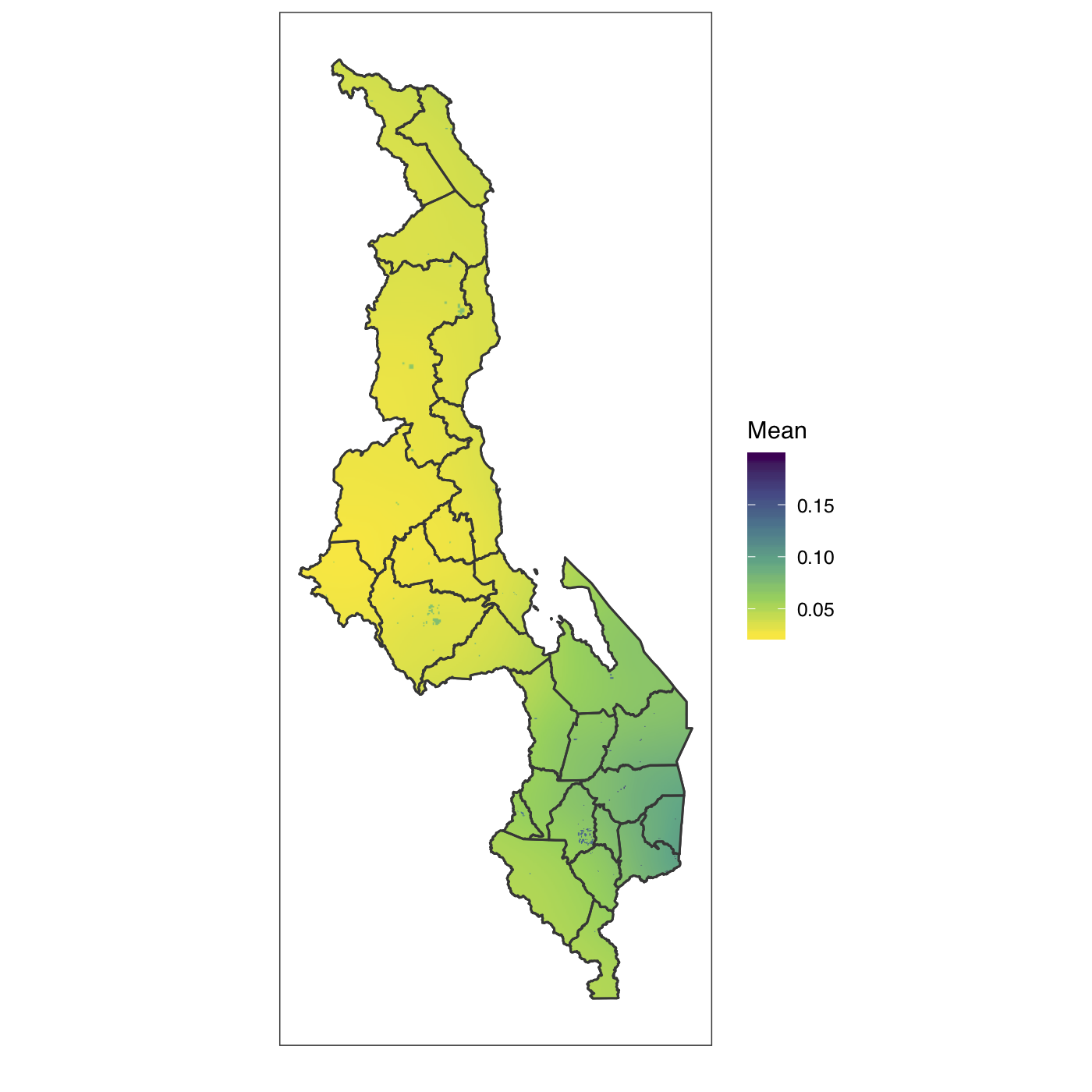}\hspace{-0.9in}
	\includegraphics[width=0.5\linewidth]{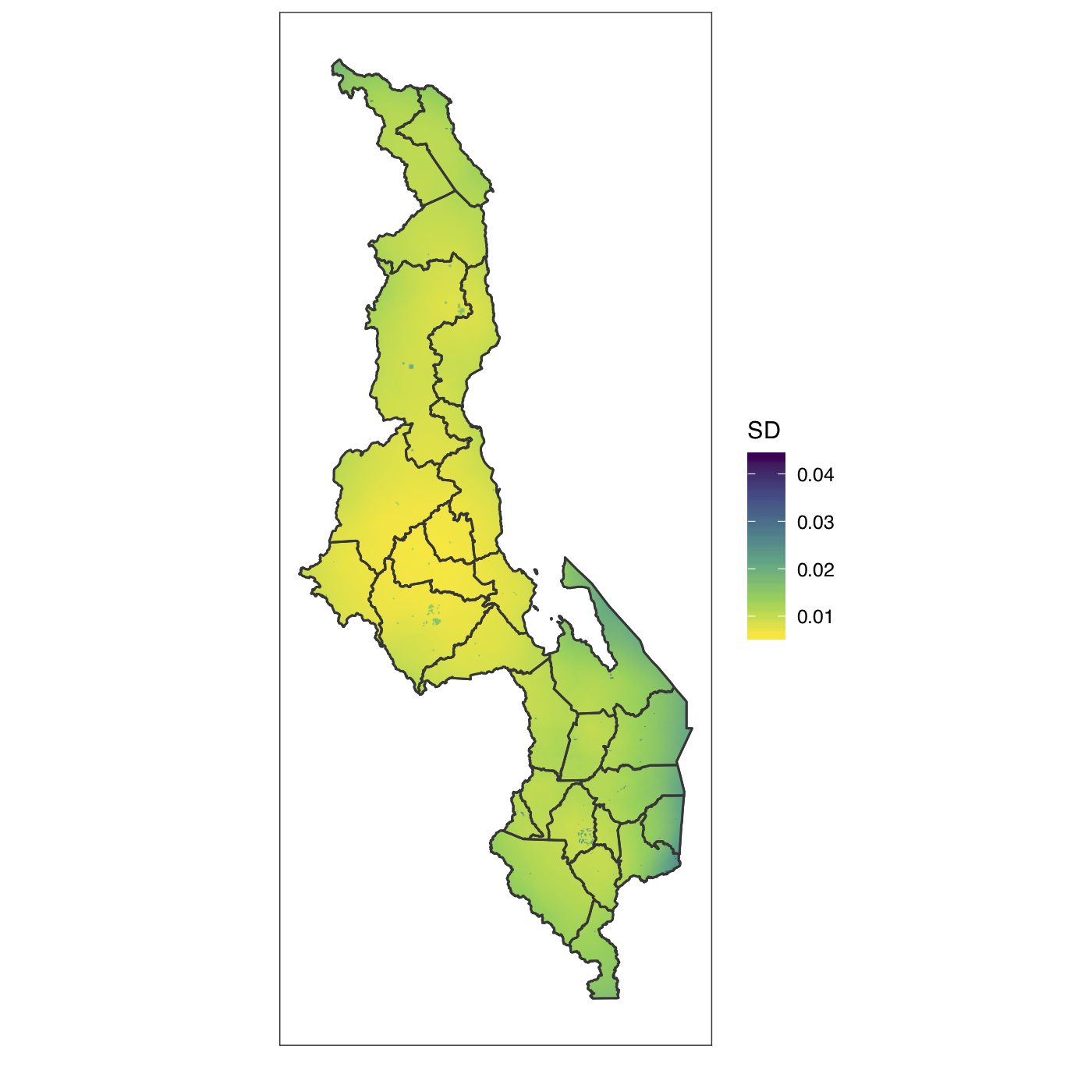}
	\caption{HIV prevalence estimates at the pixel level, using the SPDE model with no covariate. Left: point estimates (posterior mean). Right: posterior standard deviation.}
	\label{fig:SPDE2}
\end{figure}

\begin{figure}[htbp]
	\centering
	\includegraphics[width=0.5\linewidth]{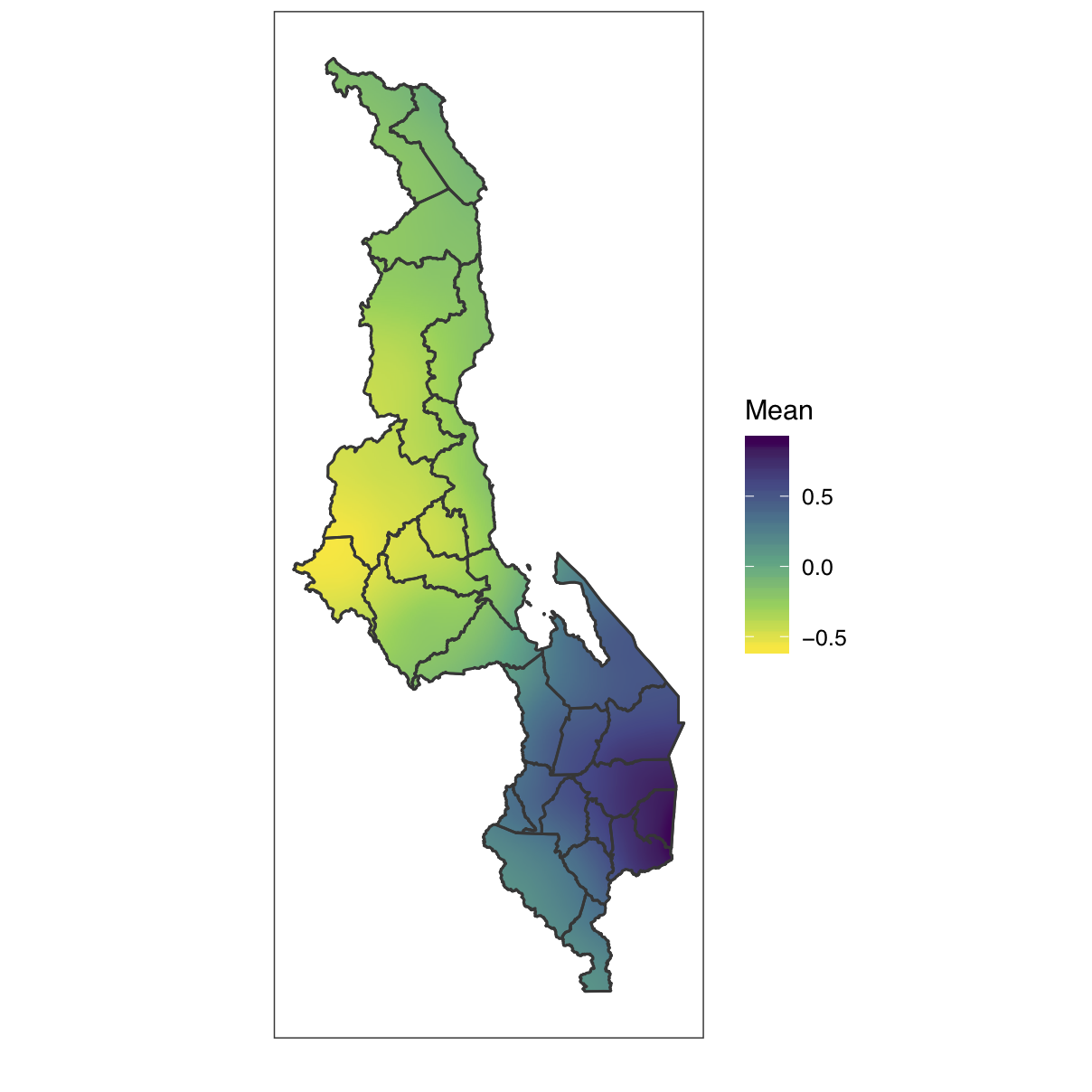}\hspace{-0.9in}
	\includegraphics[width=0.5\linewidth]{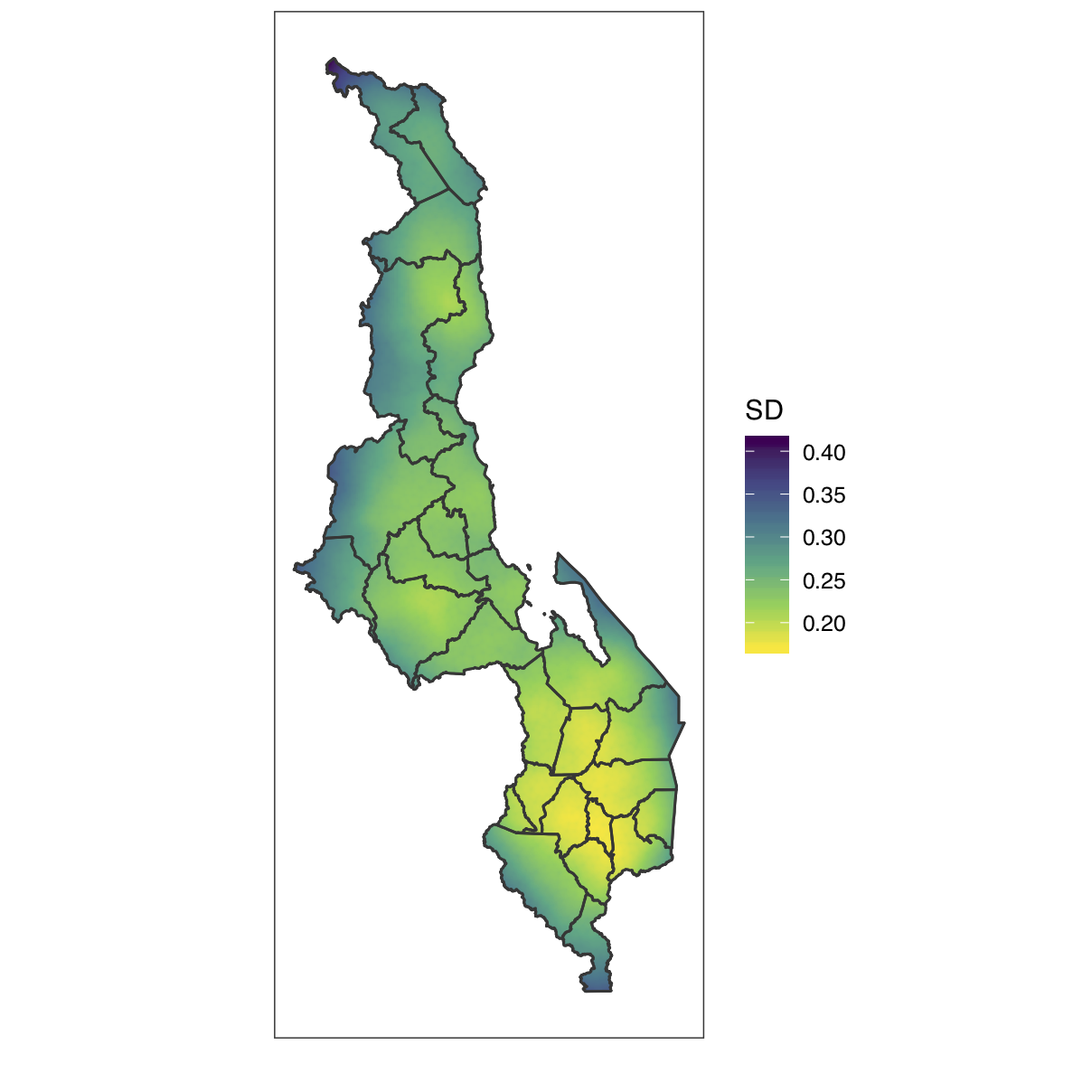}
	\caption{Spatial field at the pixel level, using the SPDE model with no covariate. Left: point estimates (posterior mean). Right: posterior standard deviation.}
	\label{fig:SPDE3}
\end{figure}

\begin{figure}[htbp]
	\centering
	\includegraphics[width=0.5\linewidth]{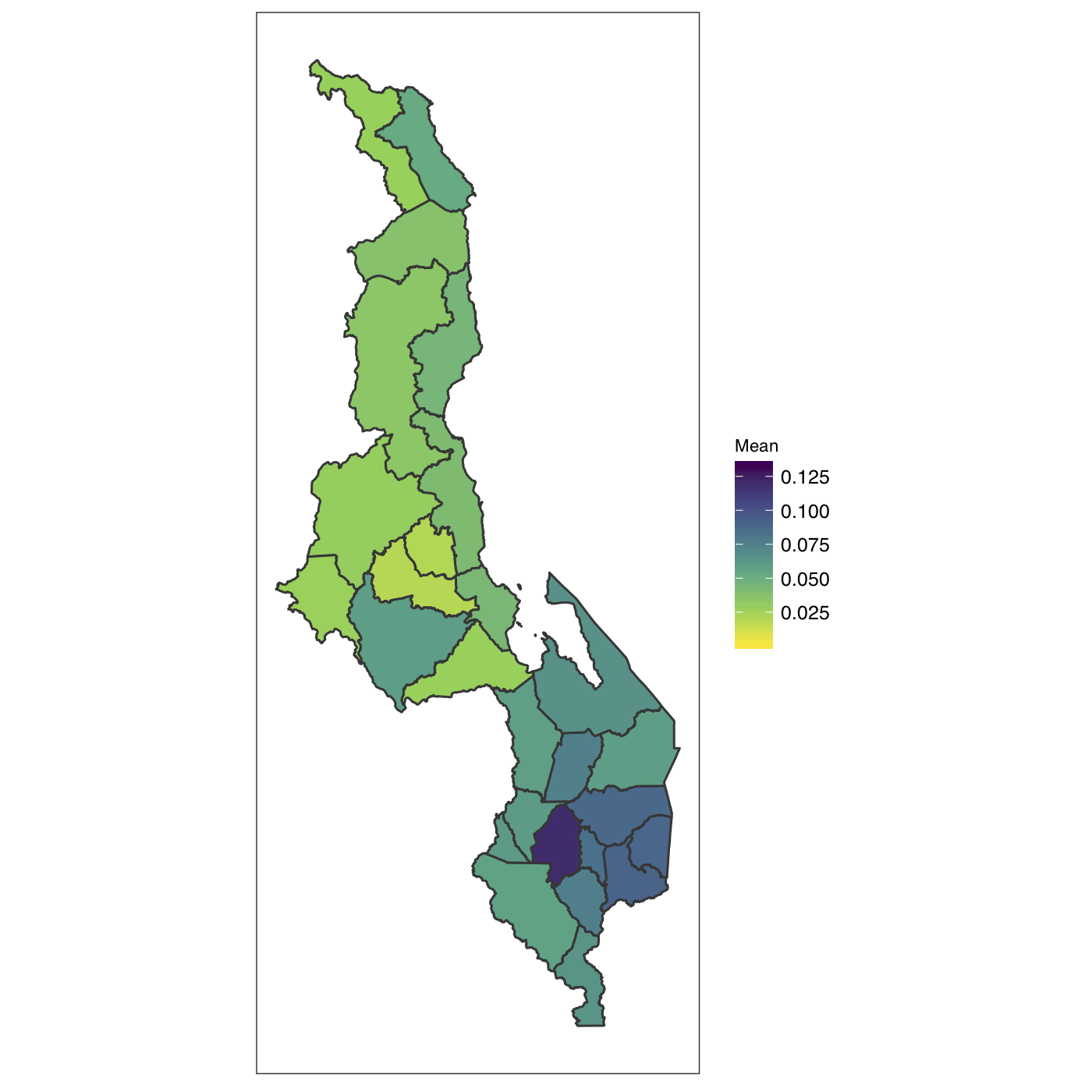}\hspace{-0.9in}
	\includegraphics[width=0.5\linewidth]{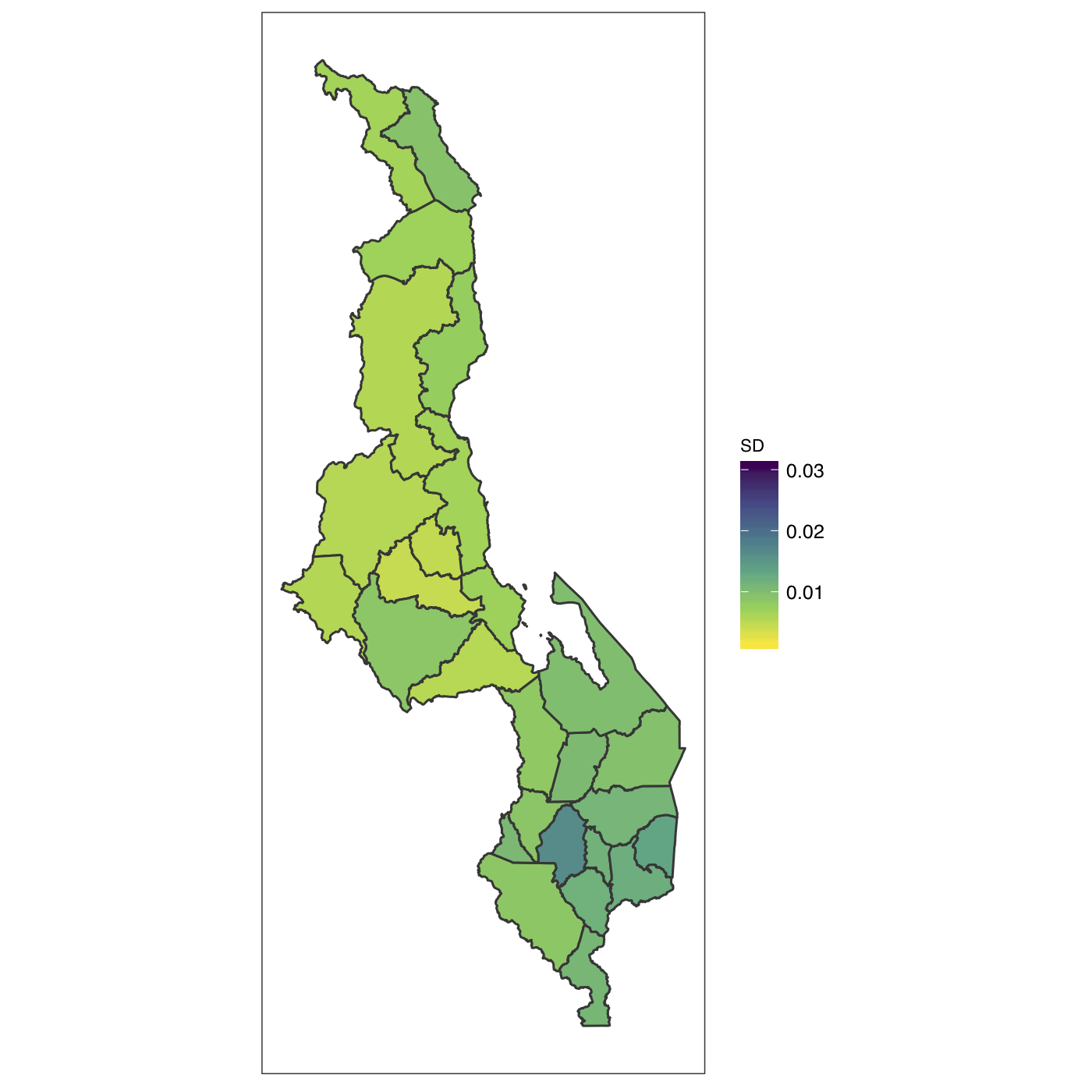}
	\caption{HIV prevalence estimates at the admin-2 level, using the SPDE model, with the ANC covariate. Left: point estimates (posterior mean). Right: posterior standard deviation.}
	\label{fig:SPDE4}
\end{figure}

\begin{figure}[htbp]
	\centering
	\includegraphics[width=0.5\linewidth]{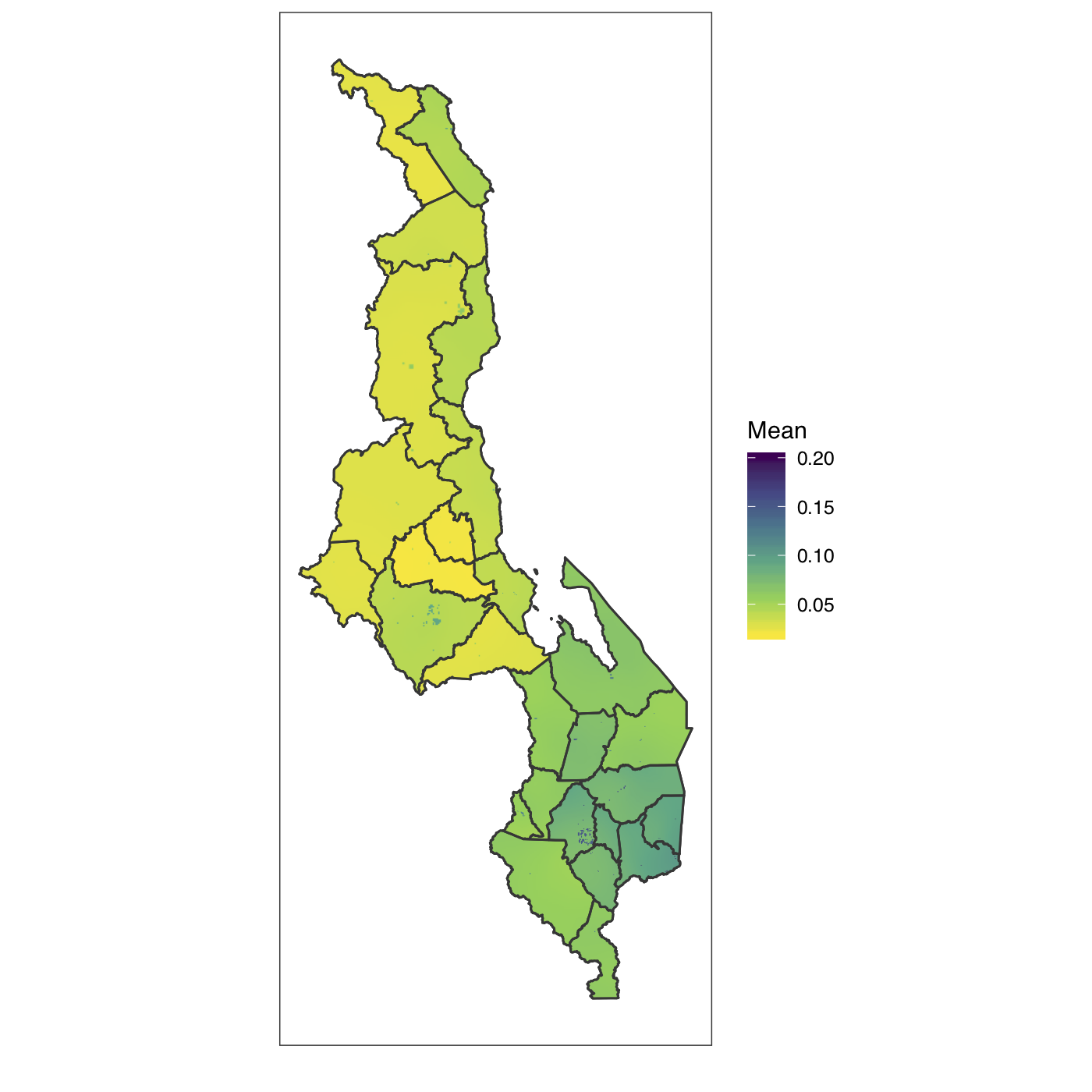}\hspace{-0.9in}
	\includegraphics[width=0.5\linewidth]{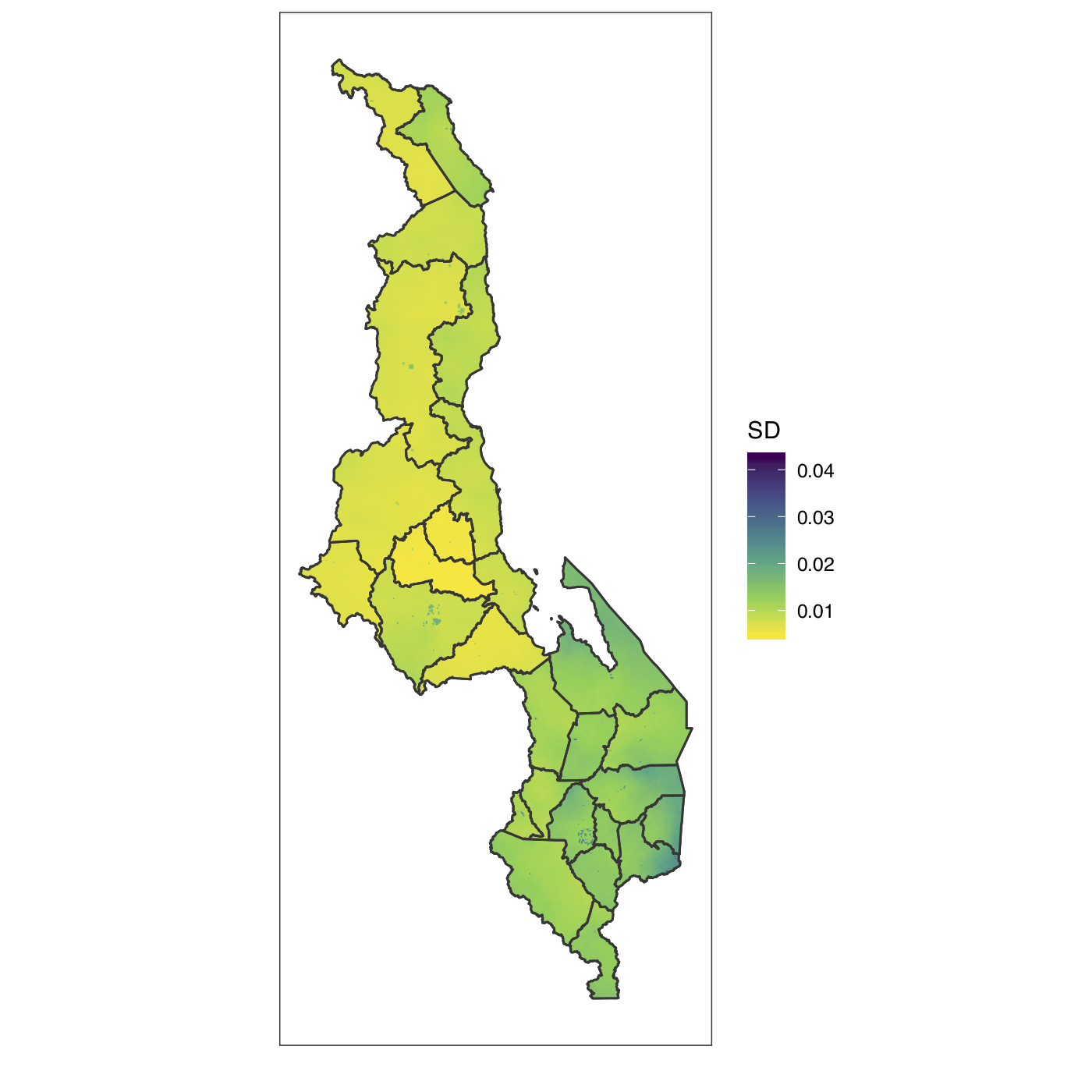}
	\caption{HIV prevalence estimates at the pixel level, using the SPDE model, with the ANC covariate. Left: point estimates (posterior mean). Right: posterior standard deviation.}
	\label{fig:SPDE5}
\end{figure}

\end{document}